\documentclass[aps,prd,showpacs,preprintnumbers,twocolumn,groupedaddress,nofootinbib]{revtex4-1}

\usepackage[]{amsmath}
\usepackage[]{amssymb}
\usepackage[]{mathrsfs}
\usepackage[]{eucal}
\usepackage[]{tensor}
\usepackage[]{amsthm}
\usepackage[]{bm}

\usepackage{tikz}

\usepackage[unicode]{hyperref}
\usepackage{xcolor}
\definecolor{pastgreen}{HTML}{669900}
\definecolor{pastblue}{HTML}{336699}
\definecolor{linkcol}{HTML}{663333}
\hypersetup{colorlinks,linkcolor={pastblue},citecolor={pastblue},urlcolor={pastblue}}

\newcommand{\pf}[1]{\mathbf{#1}}
\newcommand{\dd}{\partial}
\newcommand{\hdg}{\star} 
\newcommand{\df}{\mathrm{d}}
\newcommand{\w}{\wedge}
\newcommand{\veps}{\bm{\epsilon}}
\newcommand{\Lie}{\pounds}
\newcommand{\nab}[1]{\nabla_{\!#1}}
\newcommand{\uH}{\mathsf{H}}
\newcommand{\eH}{\buildrel H \over =}

\newcommand{\qqd}{\ , \quad}
\newcommand{\bc}{\begin{center}}
\newcommand{\ec}{\end{center}}
\newcommand{\be}{\begin{equation}}
\newcommand{\ee}{\end{equation}}

\newcommand{\defeq}{\mathrel{\mathop:}=}

\newcommand{\FF}{\mathcal{F}}
\newcommand{\GG}{\mathcal{G}}
\newcommand{\LL}{\mathscr{L}}
\newcommand{\scr}[1]{\mathrm{\scriptscriptstyle{#1}}}

\usepackage{dsfont}

\newcommand{\rr}{\mathds{R}}

\newcommand{\cl}[1]{\overline{#1}}

\theoremstyle{plain} \newtheorem{tm}{Theorem}[section]
\theoremstyle{definition} \newtheorem{defn}[tm]{Definition}
\newcommand{\btm}{\begin{tm}}
\newcommand{\etm}{\end{tm}}
\newcommand{\bdefn}{\begin{defn}}
\newcommand{\edefn}{\end{defn}}

\begin{document} 
\begin{flushleft}
\texttt{ZTF-EP-21-02}

\texttt{RBI-ThPhys-2021-6}
\end{flushleft}

\title{Black hole thermodynamics in the presence of nonlinear electromagnetic fields}

\author{A. Bokuli\'c}
\email{abokulic@phy.hr}
\affiliation{Department of Physics, Faculty of Science, University of Zagreb, 10000 Zagreb, Croatia}
\author{T. Juri\'c}
\email{tjuric@irb.hr}
\affiliation{Rudjer Bo\v skovi\'c Institute, Bijeni\v cka cesta 54, HR-10002 Zagreb, Croatia}
\author{I. Smoli\'c}
\email{ismolic@phy.hr}
\affiliation{Department of Physics, Faculty of Science, University of Zagreb, 10000 Zagreb, Croatia}

\begin{abstract}
As the interaction between the black holes and highly energetic infalling charged matter receives quantum corrections, the basic laws of black hole mechanics have to be carefully rederived. Using the covariant phase space formalism, we generalize the first law of black hole mechanics, both ``equilibrium state'' and ``physical process'' versions, in the presence of nonlinear electrodynamics fields, defined by Lagrangians depending on both quadratic electromagnetic invariants, $F_{ab}F^{ab}$ and $F_{ab}\,{\hdg F}^{ab}$. Derivation of this law demands a specific treatment of the Lagrangian parameters, similar to embedding of the cosmological constant into thermodynamic context. Furthermore, we discuss the validity of energy conditions, several complementing proofs of the zeroth law of black hole electrodynamics and some aspects of the recently generalized Smarr formula, its (non-)linearity and relation to the first law.
\end{abstract}

\maketitle

\section{Introduction} 

Thermodynamics has played a pivotal historical role in our understanding of the internal structure of matter. Establishment of the laws of black hole mechanics \cite{BCH73} and their correspondence to the basic laws of thermodynamics \cite{Bekenstein73,Bekenstein74,Hawking74} provides us with a similar invaluable guiding insight into the elusive microscopic nature of spacetime. Stationary black holes have constant surface gravity and gauge scalar potentials (zeroth law), obey energy constraints upon perturbations (first law), Hawking's law of nondecreasing horizon area (second law), and Smarr formula (Gibbs--Duhem equation). Augmented by the theoretical prediction of Hawking's radiation, there is a strong indication that the black hole surface gravity and horizon area correspond, respectively, to the temperature and entropy.

\smallskip  

Over the course of five decades vast effort has been invested into understanding of various aspects of black hole thermodynamics beyond the original Einstein--Maxwell context. Whereas far greater progress has been made in gravitational sector \cite{Wald99}, culminating in Wald's entropy formula \cite{Wald93} and its subsequent generalizations \cite{Tachikawa06,BCDPPS11}, the gauge sector still lacks a unifying picture, especially with respect to nonlinear generalizations of the classical Maxwell's electrodynamics.

\smallskip

Nonlinear electrodynamics (NLE) is an umbrella term for a broad class of theories, usually those defined by a Lagrangian constructed from two quadratic electromagnetic invariants, $F_{ab}F^{ab}$ and $F_{ab}\,{\hdg F}^{ab}$. In order to simplify nomenclature, we may sort the NLE theories into the \emph{$\FF$-class}, with Lagrangians depending only on invariant $F_{ab} F^{ab}$, and the \emph{$\FF\GG$-class}, with Lagrangians depending on both invariants. Earliest NLE theories appeared in 1930s, at the dawn of the quantum field theory. In order to cure the inconsistencies of the Maxwell's electrodynamics associated with the infinite self-energy of the point charges, Max Born proposed an $\FF$-class NLE theory \cite{Born34}, which was subsequently expanded in collaboration with Leopold Infeld to a $\FF\GG$-class NLE theory \cite{BI34}. Born--Infeld (BI) theory reappeared half a century later, at the beginning of the first superstring revolution, in low energy limits of the string theory \cite{FT85}, with the string tension $\alpha'$ and the BI parameter $b$ being related via $2\pi\alpha' = 1/b$ \cite{Tseytlin99} (for analysis on lattice see \cite{KS06}). On the other hand, not long after the work of Born and Infeld, Werner K.~Heisenberg and Hans H.~Euler \cite{HE36} found a one-loop QED correction to Maxwell's Lagrangian.

\smallskip

Nonlinearities in the electromagnetic interaction are revealed in the scattering of ``light by light'', that is the $\gamma\gamma \to \gamma\gamma$ process, and the first direct experimental evidence was recently found by the ATLAS Collaboration \cite{ATLAS17}, leading to strengthening of the constraints on parameters of the NLE Lagrangians \cite{EMY17,NAM17,NAM18} (for an overview of earlier experimental constraints on NLE theories see \cite{BR13,FBR16}). Also, complementary to the conclusions coming from experiments performed in terrestrial particle colliders, there are cosmological constraints \cite{BLM12}, as well as proposed neutrino astrophysics tests \cite{MCLP17}.

\smallskip

Interest in NLE theories within gravitational physics was ignited by realization that some modifications of the Maxwell's electrodynamics may resolve the black hole singularities, up to constraints given by \cite{BMSS79,Bronnikov00} (see also \cite{BH02,Bronnikov17}). Unfortunately, neither electrically charged Einstein--Born--Infeld black holes \cite{GSP84,SGP87,FK03,Dey04} nor electrically charged Einstein--Euler--Heisenberg black holes \cite{YT00,RWX13} are regular. Early analyses of static spherically symmetric solutions of gravitational-NLE Maxwell's equations appeared back in 1960s \cite{Peres61,PT69}, with further developments in \cite{deO94,DARG09}. A prominent example of a regular black hole spacetime, proposed by Bardeen \cite{Bardeen68}, was much later interpreted by Ay\'on-Beato and Garc\'ia \cite{ABG98,ABG00} as a solution of Einstein-NLE Maxwell field equations for a particular NLE theory (and generalized to a rotating solution in \cite{TAAS14,AA14}). Over the years the quest for a regular black hole solution became intertwined with proliferation of new NLE theories based on various Lagrangian functions, such as logarithmic \cite{Soleng95}, hyperbolic tangent \cite{ABG99}, power \cite{HM07,HM08}, exponential \cite{Hendi13}, and so on (some more recent attempts \cite{CM20,CGOP20} are based on the so-called quasitopological electromagnetism). We note in passing that Wald's solution \cite{Wald74}, describing a black hole immersed in a homogeneous magnetic field, has been recently perturbatively generalized to NLE theories \cite{BS19}.

\smallskip

The first systematic approach to thermodynamics of black holes with NLE fields by Rash\-eed \cite{Rasheed97} contains a proof of the zeroth law of black hole electrodynamics (via Einstein's gravitational field equation), an incomplete attempt to prove the first law of black hole thermodynamics (missing the crucial NLE terms), and an ambiguous conclusion that the Smarr formula does not hold. Two subsequent decades of research in this subfield brought a series of papers that mostly focused on the simplest, static spherically symmetric black hole solutions. Here we have analyses of the black hole thermodynamics for some specific theories (e.g.~power-Maxwell in arbitrary number of dimensions \cite{GHM09}, Born--Infeld \cite{CPW04,YH10,GKM12}, and Euler--Heisenberg \cite{MB20}) or more general discussions (Smarr formula via assumed first law and scaling arguments \cite{Breton04}; electrically charged black holes \cite{DARG13} but with highly implicit form of the first law and Smarr formula; Smarr formula for the $\FF$-class NLE Lagrangian, using assumed first law and scaling argument \cite{FW16}; various analyses of phase transitions in the presence of a cosmological constant and NLE fields \cite{GKM12,HV13,Azreg15,MB20}; thermodynamical stability \cite{BPB15}). Early attempt \cite{MO11a,MO11b} to generalize the first law using more rigorous, covariant phase formalism, for static black holes with constant-curvature transversal $(D-2)$-dimensional section within the $\FF$-class NLE theories, suggested the absence of NLE corrections. However, the first complete generalization of the Smarr formula for a rotating black hole with NLE fields \cite{GS17} has revealed presence of additional NLE terms, inconsistent with the unaltered form of the first law (see also remarks in \cite{MZ14}). Derivation of the first law for the $\FF$-class NLE theories \cite{ZG18}, obtained by variation of the Bardeen--Carter--Hawking mass formula, offers an important step toward the resolution of this problem.

\smallskip

The scope of this paper is broad, motivated by the fact that a proper understanding of the black hole thermodynamics in the presence of NLE fields is still quite incomplete, with numerous assumptions and technical details being usually swept under the rug. Most importantly, we shall offer complete, rigorous derivation of the first law for the rotating black holes with electromagnetic fields defined by the Lagrangian which is a member of the $\FF\GG$-class NLE theories. Necessity of such generalizations is emphasized by the fact that QED corrections to classical Maxwell's electrodynamics, defined by Euler--Heisenberg Lagrangian, is an $\FF\GG$-class NLE theory. Only when a consistent framework of black hole mechanics is reached, we can hope to distillate clear physical points and speculate about the implications of these generalizations.

\smallskip

The paper is organized as follows. In Sec.~2, we briefly review the basic elements of NLE theories, while in Sec.~3 we analyse the conditions leading to energy conditions and comment on their consequences. In Sec.~4, we revisit and complete several different, complementing approaches to the proof of the zeroth law of black hole nonlinear electrodynamics. Section 5 is the central part of the paper, where we put covariant phase space under scrutiny in order to prepare it for NLE theories, then derive the first law of black hole thermodynamics in the presence of NLE fields, both ``equilibrium state'' and ``physical process'' versions. In Sec.~6, we discuss several aspects of the NLE Smarr formula, its consistency with the first law and conditions under which it can take a linear form. In Appendices, we collect important identities, discuss Stokes' theorem on Lorentzian manifolds, and present a brief list of most important NLE Lagrangians.

\smallskip

\emph{Notation and conventions}. Throughout the paper, we use the ``mostly plus'' metric signature and the natural system of units, such that $G = c = 4\pi\epsilon_0 = 1$. Spacetime $(\mathscr{M},g_{ab})$ is a four-dimensional, connected, smooth manifold $\mathscr{M}$ with a smooth Lorentzian metric $g_{ab}$. We denote differential forms either by ``indexless'' boldface letters or with abstract index notation, whenever the former becomes cumbersome. Volume 4-form is denoted by $\veps = {\hdg 1}$. Contraction of a symmetric tensor $S_{ab}$ with a vector $X^a$ produces a 1-form $S_{ab} X^b$, which we briefly denote by $\pf{S}(X)$. Commutator between two vector fields, $X^a$ and $Y^a$ is denoted by $[X,Y]^a \defeq X^b \nab{b} Y^a - Y^b \nab{b} X^a$. On-shell equalities are denoted by $\approx$.

\section{NLE in a nutshell} 

Let us briefly review basic elements of the Einstein-NLE field equations. The NLE Lagrangian density $\LL(\FF,\GG)$ considered here is a smooth function of two electromagnetic invariants
\be
\FF \defeq F_{ab} F^{ab} \quad \textrm{and} \quad \GG \defeq F_{ab} \, {\hdg F}^{ab} \ .
\ee
For example, classical Maxwell's Lagrangian density is $\LL^\scr{(Max)} = -\FF/4$, while an overview of commonly used NLE Lagrangians is presented in Appendix \ref{app:NLEs}. It may seem that one could construct even more general NLE Lagrangians by inclusion of invariants such as $\tensor{F}{^a_b} \tensor{F}{^b_c} \tensor{F}{^c_a}$ and $\tensor{F}{^a_b} \, \tensor{{\hdg F}}{^b_c} \tensor{F}{^c_a}$. However, it is not too difficult to see, using identities (\ref{eq:FFHFHF}) and (\ref{eq:FHF}), that any scalar constructed from $\pf{F}$ and ${\hdg \pf{F}}$ without any additional derivatives, may be reduced to a function of two basic quadratic invariants $\FF$ and $\GG$ \cite{EU14}. This does not hold any more once we include, for example, covariant derivatives of electromagnetic 2-form $\pf{F}$ or nonminimal coupling to gravitation, which we will not pursue here. In order to simplify expressions, partial derivatives of the Lagrangian density $\LL$ are denoted with abbreviations such as $\LL_\FF \defeq \dd_\FF\LL$, $\LL_\GG \defeq \dd_\GG\LL$, $\LL_{\FF\GG} \defeq \dd_\GG\dd_\FF \LL$, and so on. 

\smallskip

We assume that the gravitational part of the action is the standard, Einstein--Hilbert one, so that the total Lagrangian 4-form is
\be\label{eq:Lform}
\pf{L} = \frac{1}{16\pi} \, \big( R + 4\LL(\FF,\GG) \big) \, \veps \ .
\ee
The corresponding Einstein's gravitational field equation is
\be\label{eq:Einst}
G_{ab} = 8\pi T_{ab}
\ee
with the NLE energy-momentum tensor\footnote{For a Lagrangian 4-form $\pf{L} = \varsigma \left( R + 4\LL^\scr{(em)} \right) \veps$ with normalization $\varsigma > 0$, the electromagnetic energy-momentum tensor is defined as
$$T_{ab}^{\scr{(em)}} \defeq -\frac{1}{8\pi\varsigma} \, \frac{1}{\sqrt{-g}} \, \frac{\delta S^\scr{(em)}}{\delta g^{ab}} \ , \quad \textrm{with} \quad S^\scr{(em)} = 4\varsigma \int \LL^\scr{(em)} \veps \ .$$
Our choice $\varsigma = 1/(16\pi)$ is consistent with, for example, \cite{HEll,GW01}, whereas $\varsigma = 1$ normalization is used in \cite{Wald}.}
\be
T_{ab} = -\frac{1}{4\pi} \, \big( \left( \LL_\GG \GG - \LL \right) g_{ab} + 4\LL_\FF F_{ac} \tensor{F}{_b^c} \big) \ .
\ee
The NLE Maxwell's equations are
\be\label{eq:NLEMax}
\df \pf{F} = 0 \qquad \textrm{and} \qquad \df {\hdg\pf{Z}} = 0 \ ,
\ee
where we have introduced auxiliary 2-form
\be
\pf{Z} \defeq -4 \left( \LL_\FF \pf{F} + \LL_\GG \, {\hdg\pf{F}} \right) .
\ee
We shall refer to the system of equations (\ref{eq:Einst})-(\ref{eq:NLEMax}) as the gravitational-NLE (gNLE) field equations. An alternative, convenient way to write the NLE energy-momentum tensor is to separate it into ``Maxwell part'' and the ``trace part'',
\be\label{TNLE}
T_{ab} = -4\LL_\FF T_{ab}^{\scr{(Max)}} + \frac{1}{4}\,T g_{ab} 
\ee
with
\be
T_{ab}^{\scr{(Max)}} \defeq \frac{1}{4\pi} \left( F_{ac} \tensor{F}{_b^c} - \frac{1}{4} \, g_{ab} \FF \right) 
\ee
and
\be
T \defeq g^{ab} T_{ab} = \frac{1}{\pi} \left( \LL - \LL_\FF \FF - \LL_\GG \GG \right) \ .
\ee
Note that the Maxwell's energy-momentum tensor $T_{ab}^{\scr{(Max)}}$ is traceless. Yet another way to write the NLE energy-momentum tensor, using identity (\ref{eq:FHF}), is
\be\label{TZL}
T_{ab} = \frac{1}{4\pi} \, \big( Z_{ac} \tensor{F}{_b^c} + \LL g_{ab} \big) \ .
\ee
Throughout the discussion some special spacetime points will recurringly appear as a technical obstacle. We say that an electromagnetic field is \emph{degenerate} at point $x \in \mathscr{M}$ if $\LL_\FF(x) = 0$. Whereas the Born--Infeld theory is devoid of degenerate points ($\LL^\scr{(BI)}_\FF$ does not have any real zeros), the Euler--Heisenberg theory formally has a degenerate point whenever $\FF = 45 m_e^4/4\alpha^2$, but this is inconsistent with the assumption of a weak field limit, with which this form of the Lagrangian has been written. Moreover, one might argue that at least in a weak field limit, that is near the origin of the $\FF$-$\GG$ plane, the derivative $\LL_\FF$ should take values in a neighbourhood of Maxwellian $-1/4$, without any zeros.

\section{Energy conditions} 

Measurements of macroscopic physical fields support local positivity of the energy density and its dominance over the pressure. These observations are captured by various (pointwise) energy conditions \cite{Curiel14}, among which the four most known are as follows:

\begin{itemize}
\item dominant energy condition (DEC): 

$T_{ab} u^a v^b \ge 0$ for all future directed timelike vectors $u^a$ and $v^a$ or, equivalently, $-\tensor{T}{^a_b} v^b$ is future directed causal vector for any future directed timelike vector $v^a$;

\item weak energy condition (WEC): 

$T_{ab} v^a v^b \ge 0$ for any future directed timelike vector $v^a$;

\item null energy condition (NEC): 

$T_{ab} \ell^a \ell^b \ge 0$ for any future directed null vector $\ell^a$;

\item strong energy condition (SEC): 

$T_{ab} v^a v^b \ge \frac{1}{2} \, T g_{ab} v^a v^b$ for any future directed timelike vector $v^a$.
\end{itemize} 

\noindent
Energy conditions listed above are not independent, but are related by implications 

\bc
DEC $\Rightarrow$ WEC $\Rightarrow$ NEC $\Leftarrow$ SEC. 
\ec

Foundational results in general relativity, for example, singularity theorems \cite{HEll}, are universal on the account of relying on very few details about physical fields, the most prominent being some of the energy conditions \cite{Curiel14}. As one of the versions of the zeroth law of black hole mechanics assumes that the energy-momentum tensor satisfies DEC \cite{Wald} and Hawking's black hole area law \cite{HEll,Wald} assumes that the energy-momentum tensor satisfies NEC, we shall look more closely into these conditions for NLE theories. 

\smallskip

Analysis of the energy conditions for the electromagnetic energy-momentum tensor is easiest to perform in spinorial formalism \cite{PR1,Stewart}. The electromagnetic 2-form $\pf{F}$ and its Hodge dual ${\hdg\pf{F}}$ correspond, respectfully, to spinors $F_{ABA'B'}$ and ${\hdg F}_{ABA'B'}$, which may be decomposed as
\begin{align}
F_{ABA'B'} & = \epsilon_{AB} \cl{\phi}_{A'B'} + \phi_{AB} \epsilon_{A'B'} \ , \\
{\hdg F}_{ABA'B'} & = i \left( \epsilon_{AB} \cl{\phi}_{A'B'} - \phi_{AB} \epsilon_{A'B'} \right) ,
\end{align}
with symmetric (electromagnetic) spinor $\phi_{AB}$ and antisymmetric nondegenerate spinor $\epsilon_{AB}$ (symplectic structure on spinor space)\footnote{Here we assume ``\textbf{l}eft to \textbf{l}ower, \textbf{r}ight to \textbf{r}ise'' convention of lowering and raising of spinor indices, $\epsilon_{AB} \alpha^A = \alpha_B = -\epsilon_{BA} \alpha^A$ and $\epsilon^{AB} \alpha_B = \alpha^A = -\epsilon^{BA} \alpha_B$.}. Furthermore, contraction of electromagnetic spinors admits decomposition
\be
\phi_{AC} \, \tensor{\phi}{_B^C} = \frac{1}{2}\,\epsilon_{AB} \, \phi_{DC} \phi^{DC} \ .
\ee
One must be cautious about conventions, as spinor formalism is usually done in the ``mostly minus'' metric signature. Suppose that $\eta \defeq \mathrm{sgn}(\eta_{00})$. Then the spacetime metric $g_{ab}$ corresponds to spinor $g_{ABA'B'} = \eta \epsilon_{AB} \epsilon_{A'B'}$ and
\begin{align}
\eta & F_{ACA'C'} \tensor{F}{_B^C_{B'}^{C'}} = -2\phi_{AB} \, \phi_{A'B'} + \nonumber \\
 & + \frac{1}{2} \, \epsilon_{AB} \, \epsilon_{A'B'} \left( \phi_{CD} \, \phi^{CD} + \cl{\phi}_{C'D'} \, \cl{\phi}^{C'D'} \right) \ .
\end{align}
Electromagnetic invariants are
\be
\FF = 2 \left( \phi^{AB} \phi_{AB} + \cl{\phi}^{A'B'} \cl{\phi}_{A'B'} \right)
\ee
and
\be
\GG = -2i \left( \phi^{AB} \phi_{AB} - \cl{\phi}^{A'B'} \cl{\phi}_{A'B'} \right) .
\ee
Given that we normalize Maxwell's energy-momentum tensor as
\be
T^\scr{(Max)}_{ab} \defeq -\eta\,\frac{1}{4\pi} \left( F_{ac} \tensor{F}{_b^c} - \frac{1}{4}\,g_{ab} F_{cd} F^{cd} \right) \ ,
\ee
the corresponding spinor representation reduces to
\be\label{TMaxspin}
T^\scr{(Max)}_{ABA'B'} = \frac{1}{2\pi}\,\phi_{AB} \cl{\phi}_{A'B'}
\ee
independently of the metric signature sign $\eta$. Finally, electromagnetic spinor may be decomposed \cite{PR1,Stewart} as $\phi_{AB} = \alpha_{(A} \beta_{B)}$. If $\alpha_A$ and $\beta_A$ are not proportional, then we say that $\phi_{AB}$ is algebraically general (type I in Petrov classification), whereas in case when $\alpha_A$ and $\beta_A$ are proportional, we say that $\phi_{AB}$ is algebraically special (type N). Spinor $\phi_{AB}$ is algebraically special if and only if the electromagnetic fields are null, that is, $\FF = 0 = \GG$.

\smallskip

It is well known that Maxwell's electromagnetic energy-momentum tensor (\ref{TMaxspin}) satisfies both DEC and, since it is traceless, SEC. Namely, for any pair of spinors $\kappa^A$, $\lambda^A$ and the corresponding pair of future directed null vectors, $k^{AA'} = \kappa^A \cl{\kappa}^{A'}$ and $\ell^{AA'} = \lambda^A \cl{\lambda}^{A'}$, we have
\begin{align}
T^\scr{(Max)}_{ABA'B'} k^{AA'} \ell^{BB'} & = \frac{1}{2\pi}\,\phi_{AB}\cl{\phi}_{A'B'} \kappa^A \cl{\kappa}^{A'} \lambda^B \cl{\lambda}^{B'} = \nonumber\\
 & = \frac{1}{2\pi} \left| \phi_{AB}\kappa^A\lambda^B \right|^2 \ge 0 \ .
\end{align}
Since any future directed causal vector is a sum of a pair of future directed null vectors, it follows that $T^\scr{(Max)}_{ab} u^a v^b \ge 0$ for any pair of future directed causal vectors $u^a$ and $v^a$.

\smallskip

Let us now present a simple way to translate energy conditions for NLE theories, which complements some earlier attempts \cite{Plebanski70,DDSD17}.

\smallskip

\btm\label{tm:NLEnerg}
The NLE energy-momentum tensor, in $\eta = -1$ signature, satisfies
\begin{itemize}
\item NEC if and only if $\LL_\FF \le 0$;
\item DEC if and only if $\LL_\FF \le 0$ and $T \le 0$;
\item SEC if $\LL_\FF \le 0$ and $T \ge 0$.
\end{itemize}
\etm

\emph{Proof}. One direction of the claims, the ``if'' direction, follows immediately from the (\ref{TNLE}) form of the NLE energy-momentum tensor and the fact that Maxwell's electromagnetic energy-momentum tensor $T^\scr{(Max)}_{ab}$ satisfies DEC.

For the converse in the NEC case, we need to find a future directed null vector $\ell^a$, such that $T^\scr{(Max)}_{ab}\ell^a\ell^b > 0$. Using decomposition $\phi_{AB} = \alpha_{(A} \beta_{B)}$, for the algebraically general case we may choose auxiliary spinor $\lambda^A  = \alpha^A + \beta^A$, so that $\lambda^A \alpha_A \ne 0 \ne \lambda^A \beta_A$, while in the algebraically special case $\lambda^A$ may be any spinor such that $\lambda^A \alpha_A \ne 0$. Furthermore, let $\ell^{AA'} = \pm \lambda^A \cl{\lambda}^{A'}$, with sign choice such that $\ell^a$ is future directed. Then, in both algebraically general and special cases, we have $2\pi T^\scr{(Max)}_{ab}\ell^a\ell^b = \left| \phi_{AB}\lambda^A\lambda^B \right|^2 > 0$. Finally, assuming that NEC holds, we have $0 \le T_{ab} \ell^a \ell^b = -4\LL_\FF T^\scr{(Max)}_{ab}\ell^a\ell^b$, so that $\LL_\FF \le 0$. 

Since either DEC or SEC implies NEC, given that NLE energy-momentum tensor satisfies any of these two energy conditions, it follows that $\LL_\FF \le 0$. Proof of the remaining claim, that DEC implies $T \le 0$, has already appeared in \cite{Plebanski70}, which we briefly sketch here. If $\LL_\FF = 0$, DEC immediately implies $T \le 0$, so let us assume that $\LL_\FF < 0$. Using the Newman--Penrose null tetrad \cite{Stewart}, $( \ell^a = o^A \cl{o}^{A'}, n^A = \iota^A \cl{\iota}^{A'}, m^a = o^A \cl{\iota}^{A'}, \cl{m}^a = \iota^A \cl{o}^{A'} )$, we may decompose a timelike vector $v^a$ appearing in DEC as $v^a = a \ell^a + b n^a + \cl{c} m^a + c \cl{m}^a$ with some complex numbers $(a,b,c)$, normalized for convenience with $ab = 1 + |c|^2$. One of the forms of DEC, $(\tensor{T}{^a_b} v^b)(T_{ac} v^c) \le 0$, after a straightforward but tedious calculation, is reduced to an inequality $S + (1 + 2|c|^2)\LL_\FF T \ge 0$, with some quantity $S$ independent of the parameters $(a,b,c)$. Thus, the condition $T > 0$ would lead to a contradiction as we may choose arbitrarily large $|c|$. \qed

\medskip

As we may always choose a NLE Lagrangian such that $\LL(0,0) = 0$, then, given that $\LL$ is differentiable at the origin of the $\FF$-$\GG$ plane, it follows that $T = 0$ for null electromagnetic fields. In other words, at least for null electromagnetic fields, $\LL_\FF \le 0$ is sufficient condition for both DEC and SEC. 

\smallskip

Application of Theorem \ref{tm:NLEnerg} may be illustrated with the following two most prominent NLE theories:

\begin{itemize}
\item[(a)] Born--Infeld:
\be
\LL^\scr{(BI)}_\FF = -\frac{1}{4\mathcal{W}} \qqd \pi T^\scr{(BI)} = \frac{4b^2(\mathcal{W} - 1) - \FF}{4\mathcal{W}}
\ee
with
\be
\mathcal{W} \defeq \sqrt{1 + \frac{\FF}{2b^2} - \frac{\GG^2}{16b^4}} \ .
\ee
We immediately see that $\LL^\scr{(BI)}_\FF \le 0$ and, as $2\sqrt{x-y} \le 2\sqrt{x} \le x+1$ for nonnegative $x$ and $y \le x$, we have $2\mathcal{W} \le 2 + (\FF/2b^2)$, so that $T^\scr{(BI)} \le 0$. In other words, Born--Infeld theory obeys DEC and NEC.

\item[(b)] Euler--Heisenberg:
\be
\LL^\scr{(EH)}_\FF = -\frac{1}{4} + \frac{8\alpha^2}{360 m_e^4} \, \FF \nonumber
\ee
and
\be
\pi T^\scr{(EH)} = -\frac{\alpha^2}{360 m_e^4} \left( 4\FF^2 + 7\GG^2 \right) .
\ee
We see that Euler--Heisenberg theory satisfies DEC and NEC for electromagnetic fields with $\FF \le 45 m_e^4/4\alpha^2$ (e.g.~weak field, null electromagnetic field).
\end{itemize}

\noindent
In both of these theories, SEC is satisfied for null electromagnetic fields, but this condition has to be carefully examined for non-null electromagnetic fields (see e.g.~\cite{PdLeon17}).

\section{Zeroth law(s)} 

Constancy of intensive variables over stationary black hole horizons is one of the cornerstones of the black hole thermodynamics. Just as with many other black hole theorems, the choice of the assumptions required to establish this result depends on the type of generality we strive for, whether we want it to hold for solutions with particular geometric properties of the black hole (independent of the field equations) or for solutions of some particular class of field equations (independent of particular geometric details of the spacetime).

\smallskip

The zeroth law of black hole mechanics, constancy of the surface gravity $\kappa$ over the stationary black hole horizon, can be proved as follows:

\begin{itemize}
\item[(a)] using Einstein's gravitational field equations, under the assumption that matter satisfies dominant energy condition \cite{Wald},

\item[(b)] for bifurcate Killing horizons \cite{KayWald91}, and

\item[(c)] for horizons generated by Killing vector fields which satisfy some additional geometric properties \cite{RW95}.
\end{itemize}

\noindent
The zeroth law of black hole electrodynamics, constancy of the electromagnetic scalar potentials over the stationary black hole horizon, can be established using similar techniques \cite{ISm12,ISm14}, at least for Maxwell's electromagnetic fields. Nonlinear electromagnetic fields, on the other hand, demand more careful treatment. As the analyses of the NLE zeroth law in the literature are incomplete, we shall first review various approaches.

\smallskip

Suppose that spacetime $(\mathscr{M},g_{ab})$ admits a smooth Killing vector field $\xi^a$ and the electromagnetic field $\pf{F}$ inherits the symmetry, $\Lie_\xi \pf{F} = 0$. One should bear in mind that the latter assumption is rather nontrivial, as there are known electrovac spacetimes with symmetry noninheriting electromagnetic fields \cite{MW75,BGS17,ISm18}. Symmetry inheritance of the electromagnetic fields has been extensively studied within the Maxwell's theory \cite{Woo73a,Woo73b,MW75,Coll75,RT75,WY76a,WY76b,Tod06,CDPS16} and recently analysed for NLE fields \cite{BGS17}. In general the Lie derivative $\Lie_\xi \pf{F}$ is a linear combination $a{\hdg\pf{F}} + b\pf{F}$, with $b = 0$ for Maxwell's electrodynamics, and there are various sufficient conditions implying $a = 0 = b$, which we tacitly take to be satisfied. 

\smallskip

In this context, it is convenient to introduce decomposition of $\pf{F}$ to electric and magnetic fields (differential forms) with respect to the Killing vector field $\xi^a$. First of all, we have 1-forms $\pf{E} = -i_\xi \pf{F}$ and $\pf{H} = i_\xi {\hdg\pf{Z}}$ which, as a consequence of the symmetry inheritance and NLE Maxwell equations (\ref{eq:NLEMax}), are closed,
\begin{align}
\df \pf{E} & = \left( -\Lie_\xi + i_\xi \df \right) \pf{F} = 0 \ , \\
\df \pf{H} & = \left( -\Lie_\xi + i_\xi \df \right) {\hdg\pf{Z}} = 0 \ .
\end{align}
Thus, given that a domain is simply connected, we can define on it associated scalar potentials, electric $\Phi$ and magnetic $\Psi$, via
\be
\pf{E} = -\df \Phi \quad \textrm{and} \quad \pf{H} = -\df \Psi \ .
\ee
For completeness, we may introduce two additional 1-forms, $\pf{B} = i_\xi {\hdg\pf{F}}$ and $\pf{D} = -i_\xi \pf{Z}$, with the caveat that in general $\pf{B}$ and $\pf{D}$ are not closed. These 1-forms are related by 
\begin{align}
\pf{D} & = -4 \left( \LL_\FF \, \pf{E} - \LL_\GG \, \pf{B} \right) \ , \\
\pf{H} & = -4 \left( \LL_\FF \, \pf{B} + \LL_\GG \, \pf{E} \right) \ ,
\end{align}
while electromagnetic invariants may be expressed as
\begin{align}
(\xi^a \xi_a) \, \FF & = 2 (E_a E^a - B_a B^a) \ , \label{eq:FFEB} \\
(\xi^a \xi_a) \, \GG & = -4 E_a B^a \ . \label{eq:GGEB}
\end{align}
By construction, we immediately know that scalar potentials are constant along the orbits of the Killing vector field $\xi^a$, namely, $\Lie_\xi \Phi = -i_\xi \pf{E} = 0$ and $\Lie_\xi \Psi = -i_\xi \pf{H} = 0$. The question is what can be deduced about $\Phi$ and $\Psi$ on a Killing horizon $H[\xi]$, that is a null hypersurface generated by $\xi^a$.  Given that one can prove that 
\be\label{eq:xiExiH}
\bm{\xi} \w \pf{E} \eH 0 \quad \textrm{and} \quad \bm{\xi} \w \pf{H} \eH 0 \ ,
\ee
contraction with a tangent vector $X^a \in T_p H[\xi]$ implies that $(\Lie_X \Phi) \, \bm{\xi} = 0$ and $(\Lie_X \Psi) \, \bm{\xi} = 0$. Thus, at each point where $\bm{\xi} \ne 0$, we know that $\Lie_X \Phi = 0$ and $\Lie_X \Psi = 0$, whereas at point where $\xi^a = 0$ by definition we immediately have $\df \Phi = 0$ and $\df \Psi = 0$. In conclusion, (\ref{eq:xiExiH}) imply that $\Phi$ and $\Psi$ are constant over the Killing horizon $H[\xi]$. Let us review three approaches to (\ref{eq:xiExiH}) mentioned above.

\smallskip

\begin{itemize}
\item[(a)] Gravitational field equation approach \cite{Rasheed97}. Using the identity $R_{ab} \xi^a \xi^b \eH 0$ and contraction $\pi T_{ab} \xi^a \xi^b \eH -\LL_\FF E_a E^a$, Einstein's field equation implies that the electric field $E^a$ is null at each nondegenerate point of the horizon $H[\xi]$. As $\xi^a E_a = 0$, it follows that $\bm{\xi} \w \pf{E} = 0$ at any of these points. Furthermore, (\ref{eq:FFEB}) implies that $B^a$ is null as well on $H[\xi]$, so that $\bm{\xi} \w \pf{B} \eH 0$ and, consequently, $\bm{\xi} \w \pf{H} \eH 0$. The main drawback here is that it is not quite clear how to generalize the method beyond the Einstein's gravitational field equation.

\smallskip

\item[(b)] Bifurcate horizon approach is, arguably, the simplest method. We assume that the Killing horizon $H[\xi]$ is of bifurcate type, with vanishing $\xi^a$ on bifurcation surface $\mathcal{B} \subseteq H[\xi]$. The potentials $\Phi$ and $\Psi$ are immediately constant over the bifurcation surface $\mathcal{B}$ and, as they are constant along the orbits of $\xi^a$, they are constant over each component of $H[\xi]$ connected to $\mathcal{B}$. The drawback of this approach is that a horizon does not have to be of bifurcate type, most notable counterexample being extremal black hole horizons.

\smallskip

\item[(c)] Frobenius approach \cite{ISm12,ISm14,BGS17}, in which we are relying on some additional geometric conditions. Assume that the spacetime is stationary and axially symmetric, with associated Killing vector fields, respectfully $k^a$ and $m^a$, which commute, $[k,m]^a = 0$, and satisfy Frobenius condition \cite{Lee}
\be\label{eq:Frobenius}
\pf{k} \w \pf{m} \w \df\pf{k} = 0 = \pf{k} \w \pf{m} \w \df\pf{m} \ .
\ee
Furthermore, spacetime contains Killing horizon $H[\chi]$, generated by the Killing vector field $\xi^a = \chi^a \defeq k^a + \Omega_\uH m^a$, where constant $\Omega_\uH$ is the so-called ``horizon angular velocity''. Since $k^a$ and $m^a$ are tangent to $H[\chi]$ and $\chi^a$ is normal to $H[\chi]$, it follows \cite{Heusler} that
\begin{align}
k_a k^a + \Omega_\uH \, k_b m^b & \eH 0 \ , \\
k_a m^a + \Omega_\uH \, m_b m^b & \eH 0 \ , \\
(k_a k^a) (m_a m^a) & \eH (k_a m^a)^2 \ .
\end{align}
Finally, we assume that electromagnetic field inherits both symmetries, $\Lie_k \pf{F} = 0$ and $\Lie_m \pf{F} = 0$. Applying the identity
\be
i_X \Lie_Y - i_Y \Lie_X = i_X i_Y \df - \df i_X i_Y + i_{[X,Y]} \ ,
\ee
with $X^a = k^a$ and $Y^a = m^a$ on $\pf{F}$ and ${\hdg\pf{Z}}$ it follows that $F_{ab} k^a m^b$ and ${\hdg Z}_{ab} k^a m^b$ are constant. Thus, on any connected domain of the spacetime containing the points where either $k^a$ or $m^a$ vanish (an example for the latter is the rotation axis), these constants are zero and, consequently, on each nondegenerate point of such a domain ${\hdg F}_{ab} k^a m^b = 0$. These two conditions may be rephrased as
\be
\pf{k} \w \pf{m} \w {\hdg\pf{F}} = 0 \quad \textrm{and} \quad \pf{k} \w \pf{m} \w \pf{F} = 0 \ .
\ee
Contraction with $i_m i_k$ lead us to (\ref{eq:xiExiH}) on each nondegenerate point of the horizon where $m_a m^a \ne 0$. Special points on the horizon where $m_a m^a = 0$ are usually just measure zero sets at which the rotation axis is intersecting the horizon, so that constancy of a potential over the whole horizon follows from continuity of the potential.
\end{itemize}

\smallskip

In order to repeat the strategy from (c) to a static, not necessarily axially symmetric, spacetime with associated hypersurface orthogonal\footnote{We note in passing that on any open set which is devoid of degenerate points and on which $k^a$ is hypersurface orthogonal and timelike, the NLE field cannot be null; proof is essentially same as in \cite{Tod06}.} Killing vector field $k^a$ (satisfying $\pf{k} \w \df \pf{k} = 0$) and Killing horizon $H[k]$, we would need relations of the form $\pf{k} \w {\hdg\pf{F}} = 0$ and $\pf{k} \w \pf{Z} = 0$. These, however, do not necessarily hold under given assumptions, as we may have dyonic solutions. Instead, we may treat some special subcases, defined by the additional assumptions.

\begin{itemize}
\item[(e$_1$)] ``Purely electric case'' in a sense that $\pf{B} = 0$. Then (\ref{eq:FFEB}) implies that $\pf{E}$ is again null on the horizon $H[k]$, which is enough to finalize the proof as in the approach (a) above.

\item[(e$_2$)] ``Purely electric case'' in a sense that $\pf{H} = 0$, that is, $\pf{k} \w \pf{Z} = 0$. Here $\pf{k} \w \pf{H} = 0$ implies $\LL_\GG \, \pf{k} \w \pf{E} + \LL_\FF \, \pf{k} \w \pf{B} = 0$ and contraction of $\pf{k} \w \pf{Z} = 0$ with $k^a$ implies $\LL_\FF \, \pf{k} \w \pf{E} - \LL_\GG \, \pf{k} \w \pf{B} \eH 0$. Given that $(\LL_\FF)^2 + (\LL_\GG)^2 \ne 0$, we may deduce (\ref{eq:xiExiH}).

\item[(m$_1$)] ``Purely magnetic case'' in a sense that $\pf{E} = 0$. Then (\ref{eq:FFEB}) implies that $\pf{B}$ is again null on the horizon $H[k]$, which is enough to finalize the proof as in the approach (a) above.

\item[(m$_2$)] ``Purely magnetic case'' in a sense that $\pf{D} = 0$, that is, $\pf{k} \w {\hdg\pf{Z}} = 0$. Here $\pf{k} \w \pf{D} = 0$ implies $\LL_\FF \, \pf{k} \w \pf{E} - \LL_\GG \, \pf{k} \w \pf{B} = 0$ and contraction of $\pf{k} \w {\hdg\pf{Z}} = 0$ with $k^a$ implies $\LL_\GG \, \pf{k} \w \pf{E} + \LL_\FF \, \pf{k} \w \pf{B} \eH 0$. Given that $(\LL_\FF)^2 + (\LL_\GG)^2 \ne 0$, we may deduce (\ref{eq:xiExiH}).
\end{itemize}

\smallskip

Note that for the test electromagnetic fields, weak in a sense that associated energy-momentum tensor in the gravitational field equation may be neglected, approach (a) is useless, but any of the other methods may suffice.

\section{The first law} \label{sec:first} 

The first law of black hole mechanics essentially captures energy conservation for slightly perturbed black holes. Following the nomenclature from \cite{WaldQFT}, approaches to derivation of this law may be classified as follows:

\begin{itemize}
\item[(1)] Equilibrium state version, in which we are comparing two ``nearby'' stationary black hole configurations, with two varieties:
\begin{itemize}
\item[(1a)] Original, somewhat cumbersome procedure \cite{BCH73}, in which one takes variation of the Bardeen--Carter--Hawking mass formula, and

\item[(1b)] Covariant phase space formalism \cite{LW90,Wald93,IW94,Prabhu15};
\end{itemize}

\item[(2)] Physical process version, in which we look at physical, quasistatic process of matter infalling into a black hole \cite{GW01}.
\end{itemize}

Generalization of the first law of black hole mechanics in the $\FF$-class NLE theories was recently presented in \cite{ZG18}, using approach (1a). Our aim is to extend this result for rotating black holes in the $\FF\GG$-class NLE theories, using rigorous approaches (1b) and (2). 

\smallskip

The basic assumption at the foundation of the first law is that the spacetime is a solution of gNLE equations with stationary axially symmetric, asymptotically flat metric $g_{ab}$ and a symmetry inheriting electromagnetic field $\pf{F}$. Corresponding Killing vector fields are $k^a = (\dd/\dd t)^a$, timelike at infinity, and axial $m^a = (\dd/\dd\varphi)^a$, with compact orbits. As above, we assume that $k^a$ and $m^a$ commute, $[k,m]^a = 0$, and satisfy Frobenius conditions (\ref{eq:Frobenius}). Both the equilibrium state and the physical process versions of the first law inspect Cauchy surfaces intersecting the black holes. More concretely, in the former case, the spacetime contains a bifurcate Killing horizon $H[\chi]$, a pair of null hypersurfaces generated by the null Killing vector field $\chi^a = k^a + \Omega_\uH m^a$ with constant $\Omega_\uH$ and surface gravity $\kappa$, which intersect in the so-called bifurcation surface $\mathcal{B}$, a smooth, compact, embedded 2-surface. The Killing vector field $\chi^a$ vanishes on $\mathcal{B}$. Derivation of the equilibrium state version of the first law is built on a spacelike Cauchy surface $\Sigma \subseteq \mathscr{M}$, smoothly embedded in $\mathscr{M}$ with nowhere vanishing normal, whose boundary $\dd\Sigma$ consists of an asymptotically flat end and bifurcation surface $\mathcal{B} = \Sigma \cap H[\chi]$. On the other hand, in the physical process version of the first law, we only need a portion of the Killing horizon (cut by two Cauchy surfaces), which does not need to be of the bifurcate type (accordingly, none of the Cauchy surface does not have to end in bifurcation surface).

\smallskip

For any smooth closed 2-surface $\mathcal{S}$, we define the Komar mass $M_\mathcal{S}$ and the Komar angular momentum $J_\mathcal{S}$ \cite{Komar63} with integrals
\be\label{KomarMJ}
M_\mathcal{S} \defeq -\frac{1}{8\pi} \oint_\mathcal{S} {\hdg\df\pf{k}} \qquad \textrm{and} \qquad J_\mathcal{S} \defeq \frac{1}{16\pi} \oint_\mathcal{S} {\hdg\df\pf{m}} \ .
\ee
More concretely, if $\mathcal{S}$ is a ``sphere at infinity'' $S_\infty$, that is a limit of these integrals evaluated on sphere of radius $r$ as $r \to \infty$, we use simple symbols $M \defeq M_{S_\infty}$ and $J \defeq J_{S_\infty}$. In our geometric setting, ADM definitions of mass and angular momentum \cite{Wald,Heusler} coincide with $M$ and $J$. Furthermore, we define the electric charge $Q_\mathcal{S}$ and the magnetic charge $P_\mathcal{S}$ with integrals \cite{Heusler}
\be\label{KomarQP}
Q_\mathcal{S} \defeq \frac{1}{4\pi} \oint_\mathcal{S} {\hdg\pf{Z}} \qquad \textrm{and} \qquad P_\mathcal{S} \defeq \frac{1}{4\pi} \oint_\mathcal{S} \pf{F} \ .
\ee
Again, as above, we use simple symbols $Q \defeq Q_{S_\infty}$ and $P \defeq P_{S_\infty}$ for charges evaluated at infinity. It is important to note that, given that gauge 1-form $\pf{A}$ is globally well defined, Stokes' theorem implies $P_\mathcal{S} = 0$. Thus, the magnetic charge comes with a topologically nontrivial electromagnetic field, treatment of which demands the fibre bundle tools.

\subsection{Covariant phase space scrutinized} 

Before we outline the general scheme of covariant phase space formalism, we have to address one of the crucial questions for black hole mechanics with NLE fields, the role of Lagrangian (coupling) parameters. Suppose that NLE Lagrangian is defined with a finite number of real parameters, $\{\beta_1,\dots,\beta_n\}$. Given that we treat these parameters as constants which are not varied, the result will be the first law which in general is not consistent with the generalized Smarr formula. Since the Smarr formula in the presence of NLE fields may be derived \cite{GS17} completely independently of the first law, this tension must be resolved. One of the evident options is to extend the phase space with Lagrangian parameters, so that we consider them constant within fixed spacetime (i.e.~$\nab{a} \beta_i = 0$), but analyse variations which comprise changes of parameters\footnote{The authors in \cite{HSJ17} even proposed a criterion for distinction between ``physical'' and ``redundant'' Lagrangian parameters.}. Hence, in variational procedure, the NLE Lagrangian is formally treated as a function of electromagnetic invariants and parameters, $\LL(\FF,\GG;\{\beta_i\})$. Such framework is closely related to the treatment of cosmological constant $\Lambda$ in black hole thermodynamics, leading to its identification with the pressure in $V\df p$ term \cite{KRT09,KRT10,KRT11,KMT17,HJPY17}.

\smallskip

The other possible alternative is to consider even more general framework, in which Lagrangian parameters are spacetime-dependent functions \cite{RCBKPHA17}. However, note that, using (\ref{TZL}) with identities (\ref{eq:FnabF}) and (\ref{eq:HFnabF}), we have covariant divergence
\begin{align}
4\pi \nab{a} \tensor{T}{^a_b} & = \nab{a} \left( Z^{ac} F_{bc} + \LL \tensor{\delta}{^a_b} \right) \nonumber \\
 & = (\nab{a}Z^{ac}) F_{bc} + Z^{ac} (\df F)_{abc} + \sum_{i=1}^n \LL_{\beta_i} \nab{b} \beta_i \ ,
\end{align}
which for nonconstant parameters $\beta_i$ will not necessarily vanish on-shell. This indicates that one needs to complete such theory with additional equations for parameters, but we will not pursue such generalizations here.

\smallskip

We now turn to application of the covariant phase space formalism under the assumptions given above. In this subsection, for simplicity, we shall denote all dynamical fields (first of all, spacetime metric $g_{ab}$ and gauge field $\pf{A}$) collectively by $\phi$, with all indices suppressed. Similarly, the index of coupling parameters $\beta_i$ will be  suppressed in arguments, but we shall keep them in sums involving variations $\delta\beta_i$. Within the variational procedure, we assume that the action of the ``variation operator'' $\delta$ on fields $\phi$ and parameters $\beta_i$ is defined \cite{LW90,Wald} as
\be
\delta\phi(x) \defeq \frac{\dd\phi(x;\lambda)}{\dd\lambda} \Big|_{\lambda=0} \quad \textrm{and} \quad \delta\beta_i \defeq \frac{\dd\beta_i(\lambda)}{\dd\lambda} \Big|_{\lambda=0} \ ,
\ee
where $\phi(x;\lambda)$ and $\beta_i(\lambda)$ are smooth 1-parameter configurations of fields and coupling parameters. One must bear in mind that variations of the metric and its inverse are related by
\be
\delta g_{ab} = -g_{ac} g_{bd} \, \delta g^{cd} \ ,
\ee
while the variation of the volume form may be decomposed as
\be
\delta\veps = -\frac{1}{2}\,\veps \, g_{ab} \, \delta g^{ab} \ .
\ee

Variation of the Lagrangian 4-form consists of the following terms \cite{IW94}:
\be
\delta \pf{L}[\phi;\beta] = \pf{E}[\phi;\beta] \, \delta\phi + \bm{\Lambda}^i[\phi;\beta] \, \delta\beta_i + \df\bm{\Theta}[\phi,\delta\phi;\beta] \ .
\ee
Field equations are contained in the 4-form $\pf{E}$, indexed 4-form $\bm{\Lambda}^i$ is the variation of the Lagrangian with respect to coupling parameter $\beta_i$, while the boundary terms are collected in the 3-form $\bm{\Theta}$. Next, we introduce the Noether current 3-form 
\be\label{eq:J}
\pf{J}_\xi \defeq \bm{\Theta}[\phi,\Lie_\xi \phi;\beta] - i_\xi \pf{L}[\phi;\beta] \ ,
\ee
defined with respect to an arbitrary fixed vector field $\xi^a$, which will later be promoted to a Killing vector field. Now, as
\be
\df\pf{J}_\xi = -\pf{E}[\phi;\beta] \, \Lie_\xi \phi - \bm{\Lambda}^i[\phi;\beta] \, \Lie_\xi \beta_i
\ee
and $\Lie_\xi \beta_i = 0$, the Noether 3-form is closed on-shell, $\df\pf{J}_\xi \approx 0$, and at least locally exists \cite{Wald90} a 2-form $\pf{Q}_\xi$, such that $\pf{J}_\xi \approx \df\pf{Q}_\xi$. In other words, as will be explicitly shown below, we may write
\be\label{eq:JCQ}
\pf{J}_\xi = i_\xi\pf{C} + \df\pf{Q}_\xi \ ,
\ee
where $\pf{C}$ is a 4-form, which vanishes on-shell, $\pf{C} \approx 0$. As our focus is mainly on theories with the Lagrangian which is a sum of the gravitational and electromagnetic parts, it follows that the 3-form $\bm{\Theta}$ and the 2-form $\pf{Q}_\xi$ split accordingly,
\be
\bm{\Theta} = \bm{\Theta}^\scr{(g)} + \bm{\Theta}^\scr{(em)} \quad \textrm{and} \quad \pf{Q}_\xi = \pf{Q}^\scr{(g)}_\xi + \pf{Q}^\scr{(em)}_\xi \ .\nonumber
\ee
The symplectic current 3-form is defined with respect to two variations $\delta_1$ and $\delta_2$,
\be
\bm{\omega}[\phi,\delta_1\phi,\delta_2\phi;\beta] \defeq \delta_1 \bm{\Theta}[\phi,\delta_2\phi;\beta] - \delta_2 \bm{\Theta}[\phi,\delta_1\phi;\beta] \ ,
\ee
and the presymplectic form is obtained by integrating symplectic current 3-form over a spacelike Cauchy surface $\Sigma$
\be
\Omega_\Sigma[\phi,\delta_1\phi,\delta_2\phi;\beta] \defeq \int_\Sigma \bm{\omega}[\phi,\delta_1\phi,\delta_2\phi;\beta] \ .
\ee
A tacit assumption here is that volume form (orientation) on $\Sigma$ is given by pullback of $i_{\tilde{n}} \veps$, where $\tilde{n}^a$ is a unit, future directed timelike normal on $\Sigma$. Taking into account that $\delta\xi^a = 0$, variation of the Noether current (\ref{eq:JCQ}) gives $\delta \pf{J}_\xi = i_\xi \delta\pf{C} + \df\delta\pf{Q}_\xi$, while variation of (\ref{eq:J}) leads to
\begin{align}
\delta \pf{J}_\xi & = -i_\xi \pf{E}[\phi;\beta] \delta\phi + \bm{\omega}[\phi,\delta\phi,\Lie_\xi\phi;\beta] + \nonumber\\
 & + \df i_\xi \bm{\Theta}[\phi,\delta\phi;\beta] - i_\xi \bm{\Lambda}^i[\phi;\beta] \, \delta\beta_i \ ,
\end{align}
so that
\begin{align}
\bm{\omega}[\phi, & \, \delta\phi,\Lie_\xi\phi;\beta] = i_\xi (\pf{E}\,\delta\phi + \delta\pf{C}) + \nonumber\\
 & + \df(\delta\pf{Q}_\xi - i_\xi \bm{\Theta}[\phi,\delta\phi;\beta]) + i_\xi \bm{\Lambda}^i[\phi;\beta] \, \delta\beta_i \ .
\end{align}
Immediately, using Stokes' theorem (\ref{Stokes}), we have
\begin{align}\label{Omega}
\Omega_\Sigma[\phi, & \, \delta\phi,\Lie_\xi\phi;\beta] = \int_\Sigma i_\xi (\pf{E}\,\delta\phi + \delta\pf{C}) + \nonumber\\
 & + \int_{\dd\Sigma} \left( \delta\pf{Q}_\xi - i_\xi \bm{\Theta}[\phi,\delta\phi;\beta] \right) - K_\xi^i(\beta) \, \delta\beta_i \ ,
\end{align}
where we have introduced auxiliary functions $K_\xi^i$,
\be
K_\xi^i(\beta) \defeq -\int_\Sigma i_\xi \bm{\Lambda}^i[\phi;\beta] \ .
\ee
As the top compactly supported de Rham cohomology group for smooth oriented (compact and noncompact) manifolds with nonempty boundary is trivial (see e.g.~Theorems 8.3.10 and 8.4.8 in \cite{Weintraub}), we know that pullback of the $i_\xi \bm{\Lambda}^i$ to $\Sigma$ is globally exact at least for compactly supported fields, and in this case and we can rewrite, via Stokes' theorem (\ref{Stokes}), $K_\xi^i$ as an integral over $\dd\Sigma$. On the other hand, for noncompact $\Sigma$ with fields which decay at infinity, but are not necessarily compactly supported, the problem of rewriting of $K_\xi^i$ as a boundary integral depends on further details of the theory.

\smallskip

In order to connect this procedure with Hamiltonian mechanics\footnote{
Let us do a brief recap of Hamiltonian mechanics: building elements consist of a phase space manifold with local canonical coordinates $s^\mu = (q^1,\dots,p_1,\dots)$, symplectic (closed, nondegenerate) 2-form $\bm{\omega}$ and correspondence between a function $f$ and tangent vector $X_f$ via $\df f = - i_{X_f} \bm{\omega}$, that is,
$$X_f = \frac{\dd f}{\dd p_i} \, \frac{\dd}{\dd q^i} - \frac{\dd f}{\dd q^i} \, \frac{\dd}{\dd p_i} \ .$$
Dynamics is defined by Hamiltonian scalar $H$, $\dot{f} = X_H(f)$ and variation $\delta H = (\nab{\mu} H) \delta s^\mu = \omega_{\mu\nu} \delta s^\mu \dot{s}^\nu \nonumber$.
}, encapsulated in relation $\delta H_\xi = \Omega_\Sigma[\phi,\delta\phi,\Lie_\xi\phi;\beta]$, one has to prove the existence of Hamiltonian $H_\xi$, conjugate to $\xi^a$ on $\Sigma$. Given that $\phi$ is a solution of field equations (thus $\pf{E} = 0$) and $\delta\phi$ is a solution of linearized equations (thus $\delta\pf{C} = 0$), the first integral on the right-hand side of (\ref{Omega}) will be zero. Thus, the question is whether remaining terms can be written on-shell as a variation of something.

In the absence of contribution from parameters, $K_\xi^i \, \delta\beta_i$, Hamiltonian exists \cite{WZ00} if and only if
\be
\int_{\dd\Sigma} i_\xi \bm{\omega}[\phi,\delta_1\phi,\delta_2\phi] = 0
\ee
for any two variations $\delta_1$ and $\delta_2$. More concretely, it is known \cite{IW94} that Einstein--Hilbert gravitational contribution to $i_\xi \bm{\Theta}$ term may be written as a variation, with the help of a 3-form $\pf{b}$ such that
\be
\int_{\dd\Sigma} i_\xi \bm{\Theta}^\scr{(g)} = \delta\!\int_{\dd\Sigma} i_\xi \pf{b} \ .
\ee
As will be demonstrated in the following subsection, electromagnetic contribution to $i_\xi \bm{\Theta}$ term will vanish due to boundary conditions and gauge choices. Finally, we have to address integrability of the term $K_\xi^i \, \delta\beta_i$. As local condition $\dd_{\beta_i} K_\xi^j = \dd_{\beta_j} K_\xi^i$ is satisfied under mild smoothness assumptions, we know that $I_\xi(\beta)$ exists, such that $\delta I_\xi = K_\xi^i \, \delta\beta_i$. In the simplest case, with a single coupling parameter ($n=1$), $I_\xi$ is simply a primitive function of $K_\xi$.

\smallskip

Now we specialize to a geometric setting described in the introduction of Sec.~5. First, we assume that $\xi^a$ is a Killing vector field and all dynamical fields inherit corresponding symmetry, $\Lie_\xi \phi = 0$, so that\footnote{In a more general context, this implication demands a careful justification \cite{IW94}, but here it will be immediately clear that symplectic current 3-form $\bm{\omega}[\phi,\delta\phi,\Lie_\xi\phi;\beta]$, constructed from the Einstein--Hilbert gravitational 3-form $\bm{\Theta}^\scr{(g)}$ and NLE 3-form $\bm{\Theta}^\scr{(em)}$, vanishes for Killing vector field $\xi^a$ and symmetry inheriting electromagnetic fields.} $\Omega_\Sigma[\phi,\delta\phi,\Lie_\xi\phi;\beta] = 0$. Then (\ref{Omega}) decomposes on-shell as
\be\label{eq:QbQbK}
\delta\!\oint_{S_\infty} (\pf{Q}_\xi - i_\xi\pf{b}) - \delta\!\oint_\mathcal{B} (\pf{Q}_\xi - i_\xi\pf{b}) - K_\xi^i \, \delta\beta_i \approx 0 \ .
\ee
Second, we assume that $\xi^a = \chi^a = k^a + \Omega_\uH m^a$ and identify various contributions to boundary integrals. 

\smallskip

The gravitational part of the Lagrangian 4-form (\ref{eq:Lform}) is conventional Einstein--Hilbert term, whose variational properties are well known \cite{Wald,Wald93},
\be
\frac{1}{16\pi} \, \delta (R\veps) = \frac{1}{16\pi} \, G_{ab} \, \delta g^{ab} \veps + \df\bm{\Theta}^\scr{(g)} \qqd \bm{\Theta}^\scr{(g)} \defeq \frac{1}{16\pi} \, {\hdg\bm{v}}
\ee
where $\bm{v}$ is an auxiliary 1-form defined by
\be
v_a \defeq \nabla^b \delta g_{ab} - g^{cd} \nab{a} \delta g_{cd} \ ,
\ee
and
\be
\pf{Q}_\xi^\scr{(g)} = -\frac{1}{16\pi} \, {\hdg\df\bm{\xi}} \ .
\ee
Gravitational contributions to (\ref{eq:QbQbK}) give us \cite{IW94} mass and angular momentum of the black hole spacetime, defined respectfully by
\be\label{MQSinfty}
M = \oint_{S_\infty} (\pf{Q}^\scr{(g)}_k - i_k\pf{b}) \qquad \textrm{and} \qquad J = -\oint_{S_\infty} \pf{Q}^\scr{(g)}_m \ .
\ee
The absence of the $i_m\pf{b}$ term in the integral for the angular momentum (pullback of $i_m\pf{b}$ to any sphere to which $m^a$ is tangent vanishes) is reflected in different normalization of Komar mass and angular momentum \cite{IW94}. Gravitational contribution at horizon produces the entropy term \cite{Wald93}
\be
\delta\!\oint_\mathcal{B} \pf{Q}^\scr{(g)}_\xi = \frac{\kappa}{8\pi} \, \delta \mathcal{A} \ ,
\ee
where $\mathcal{A}$ is the area of the bifurcation surface $\mathcal{B}$. Altogether, the interim form of the first law we have obtained reads
\be\label{first.inter}
\delta M - \Omega_\uH \, \delta J + \delta\!\oint_{S_\infty} \pf{Q}^\scr{(em)}_\chi = \frac{\kappa}{8\pi} \, \delta \mathcal{A} + \delta\!\oint_\mathcal{B} \pf{Q}^\scr{(em)}_\chi + K_\chi^i \, \delta\beta_i \ .
\ee

\subsection{Equilibrium state first law} 

Now we turn to the NLE contributions to the first law of black hole mechanics. Variation of the NLE Lagrangian,
\be
\delta (\LL\veps) = \LL_\FF \, \delta\FF \, \veps + \LL_\GG \, \delta\GG \, \veps + \LL \, \delta\veps + \sum_{i=1}^n \LL_{\beta_i} \, \delta\beta_i \, \veps
\ee
may be conveniently written as
\be
\delta (\LL\veps) = \LL_\FF \, \delta(\FF\veps) + \LL_\GG \, \delta(\GG\veps) + \pi T \delta\veps + \sum_{i=1}^n \LL_{\beta_i} \, \delta\beta_i \, \veps \ . \label{eq:varLeps}
\ee
The first term in (\ref{eq:varLeps}) is, up to factor $\LL_\FF$, just the standard Maxwellian contribution
\begin{align}\label{eq:var1}
\LL_\FF \, \delta(\FF\veps) & = 8\pi \LL_\FF T^\scr{(Max)}_{ab} \delta g^{ab} \veps - 4 \LL_\FF \nab{a} F^{ab} \delta A_b \veps + \nonumber\\
 & + 4 \LL_\FF \nabla^a (\tensor{F}{_a^b} \delta A_b) \veps \ .
\end{align}
Combination of the first term in (\ref{eq:var1}) and the third term in (\ref{eq:varLeps}) gives us the NLE energy-momentum tensor
\be
8\pi \LL_\FF T^\scr{(Max)}_{ab} \delta g^{ab} \veps + \pi T \, \delta\veps = -2\pi T_{ab} \delta g^{ab} \veps \ .
\ee
Also, since
\begin{align}
 & - \LL_\FF \nab{a} F^{ab} \delta A_b + \LL_\FF \nab{a} (F^{ab} \delta A_b) = \nonumber\\
 & = -\nab{a} (\LL_\FF F^{ab}) \delta A_b + \nabla^a (\LL_\FF \tensor{F}{_a^b} \delta A_b) \ , 
\end{align}
the sum of the first and the third terms in (\ref{eq:varLeps}) gives us
\begin{align}
\LL_\FF \, & \delta(\FF\veps) + \pi T \delta\veps = -2\pi T_{ab} \delta g^{ab} \veps - \nonumber\\
 & - 4 \nab{a} (\LL_\FF F^{ab}) \delta A_b \veps + 4 \nabla^a (\LL_\FF \tensor{F}{_a^b} \delta A_b) \veps \ .
\end{align}
The second term in (\ref{eq:varLeps}) may be written, using (\ref{eq:FwF}), as
\begin{align}
\LL_\GG \delta(\GG\veps) & = 4 \LL_\GG \left( \nab{a} (({\hdg F}^{ab}) \, \delta A_b) - (\nab{a}{\hdg F}^{ab}) \delta A_b \right) \veps \nonumber \\
 & = 4 \left( \nabla^a (\LL_\GG \, (\tensor{{\hdg F}}{_a^b}) \, \delta A_b) - \nab{a} (\LL_\GG \, {\hdg F}^{ab}) \delta A_b \right) \veps \ .
\end{align}
Altogether, we have obtained a sought form of the variation of the Lagrangian 4-form,
\begin{align}
\frac{1}{4\pi} \, \delta (\LL\veps) & = \frac{1}{16\pi} \Big( -8\pi T_{ab} \, \delta g^{ab} + 4 (\nab{a} Z^{ab}) \delta A_b + \nonumber\\
 & + 4 \sum_i \LL_{\beta_i} \, \delta\beta_i \Big) \veps + \df\bm{\Theta}^\scr{(em)}
\end{align}
with
\be
\bm{\Theta}^\scr{(em)} \defeq \frac{1}{16\pi} \, {\hdg\bm{w}} \qqd w_a = -4\tensor{Z}{_a^b} \, \delta A_b \ .
\ee
Auxiliary 1-form $\bm{w}$ may be also written as $\bm{w} = -4\,{\hdg({\hdg\pf{Z} \w \delta\pf{A}})}$. Here we can see \cite{Prabhu15} that for the electromagnetic field $\pf{F}$ of class $O(r^{-2})$ and perturbation $\delta\pf{A}$ of class $O(r^{-1})$ as $r \to \infty$, the 3-form $\bm{\Theta}^\scr{(em)}$ does not contribute to the integral at $S_\infty$.

\smallskip

Let us turn to Noether current 3-form
\be
16\pi \pf{J}_\xi = {\hdg(\bm{v} + \bm{w})} - \left( R + 4\LL \right) {\hdg\bm{\xi}} \ .
\ee
Using the identity
\be
\nabla^b \nab{b} \xi_a - \nab{a} \nabla^b \xi_b = \pf{R}(\xi)_a - ({\hdg\df{\hdg\df \bm{\xi}}})_a
\ee
we see that auxiliary 1-form $\bm{v}$ for $\delta = \Lie_\xi$ is equal to
\be
\nabla^b \Lie_\xi g_{ab} - g^{cd} \nab{a} \Lie_\xi g_{cd} = 2\pf{R}(\xi)_a - ({\hdg\df{\hdg\df \bm{\xi}}})_a \ .
\ee
For the NLE 1-form $\bm{w}$, we have to find objects containing contraction $\tensor{Z}{_a^b} \Lie_\xi A_b$. As the Lie derivative $\Lie_\xi \pf{A}$ is contained in the electric 1-form defined with respect to the vector field $\xi^a$,
\be
\pf{E} = -i_\xi \pf{F} = -i_\xi \df\pf{A} = -\Lie_\xi \pf{A} + \df i_\xi \pf{A}
\ee
our focus is on the contraction $i_E \pf{Z}$. Here we need one auxiliary identity,
\be
i_E\,{\hdg\pf{F}} = \frac{1}{4} \, \GG \, \bm{\xi} \ ,
\ee
following directly from (\ref{eq:FHF}), so that
\be
4 i_E \pf{Z} = -16 \left( \LL_\FF \, i_E \pf{F} + \LL_\GG \, i_E {\hdg\pf{F}} \right) = 16\pi \pf{T}(\xi) - 4\LL\bm{\xi} \ .
\ee
On the other hand,
\begin{align}
4 i_E \pf{Z} & = -4 {\hdg(\hdg\pf{Z} \w \pf{E})} \nonumber \\
 & = 4{\hdg({\hdg\pf{Z}} \w \Lie_\xi \pf{A}) - 4{\hdg({\hdg\pf{Z}} \w \df i_\xi \pf{A})}} \nonumber \\
 & = -\bm{w} - 4{\hdg\df((i_\xi \pf{A}){\hdg\pf{Z}})} + 4(i_\xi \pf{A}) {\hdg\df{\hdg\pf{Z}}}
\end{align}
which leads to
\be
\bm{w} - 4\LL \bm{\xi} = -16\pi \pf{T}(\xi) - 4{\hdg\df((i_\xi \pf{A}){\hdg\pf{Z}})} + 4(i_\xi \pf{A})\,{\hdg\df{\hdg\pf{Z}}} \ .
\ee
As the variational procedure introduces electromagnetic field via gauge 1-form $\pf{A}$, we must establish the relation between $\pf{A}$ and scalar potential. Supposing that the electromagnetic field inherits the symmetry, $\Lie_\xi \pf{F} = 0$, and $\pf{F} = \df\pf{A}_0$ for some initial gauge choice of gauge 1-form $\pf{A}_0$, we still might have a technical problem as $\Lie_\xi \pf{A}_0 \ne 0$. Then, as $\df\Lie_\xi\pf{A}_0 = \Lie_\xi\pf{F} = 0$, we know that $\Lie_\xi\pf{A}_0$ is a closed form and on a simply connected domain there is a function $\alpha$, such that $\Lie_\xi\pf{A}_0 = \df\alpha$. Thus, choosing a gauge function $\lambda$ defined by $\Lie_\xi\lambda = -\alpha$, we have $\pf{A} = \pf{A}_0 + \df\lambda$, for which $\Lie_\xi \pf{A} = 0$. Even after this gauge fixing, we still have a remaining symmetry-consistent gauge freedom, as for any function $\mu$ such that $\Lie_\xi \mu = 0$, we have $\Lie_\xi (\pf{A} + \df\mu) = 0$. Furthermore,
\be
\df (\Phi + i_\xi \pf{A}) = -\pf{E} + (\Lie_\xi - i_\xi \df) \pf{A} = 0
\ee
proves that $\Phi$ and $-i_\xi\pf{A}$ differ by a constant, say $\Phi = -i_\xi\pf{A} + \Phi_0$ for some $\Phi_0 \in \rr$. This allows us to write
\be
\pf{J}_\xi = \frac{1}{8\pi} \, {\hdg(\pf{G}(\xi) - 8\pi \pf{T}(\xi))} - \frac{\Phi - \Phi_0}{4\pi}\,\df{\hdg\pf{Z}} + \df(\pf{Q}_\xi^\scr{(g)} + \pf{Q}_\xi^\scr{(em)})
\ee
with
\be
\pf{Q}_\xi^\scr{(g)} = -\frac{1}{16\pi}\,{\hdg\df\bm{\xi}} \quad \textrm{and} \quad \pf{Q}_\xi^\scr{(em)} = \frac{1}{4\pi} \, (\Phi - \Phi_0)\,{\hdg\pf{Z}} \ .
\ee
The 4-form $\pf{C}$ is given by
\be\label{C}
C_{abcd} = \frac{1}{8\pi} \, (\tensor{G}{_a^e} - 8\pi\tensor{T}{_a^e} - 2 A_a \nab{r} Z^{re}) \epsilon_{ebcd} \ .
\ee
Again, this confirms that $\df \pf{J}_\xi \approx 0$ and $\pf{J}_\xi \approx \df\pf{Q}_\xi$.

\smallskip

Before we evaluate remaining terms for the first law (\ref{first.inter}), it is convenient to make a gauge choice. If we take $\pf{A}$ such that $i_\xi \pf{A}$ will give nonvanishing contribution at bifurcation surface, we are tacitly using gauge field which is divergent there. Take for a simple example nonextremal Reissner--Norstr\"om black hole solution. Using tortoise radial coordinate $\df r_* = \df r/f(r)$, we can introduce Eddington--Finkelstein coordinates $u = t - r_*$ and $v = t + r_*$, and then Kruskal coordinates $U = -e^{-\kappa u}$ and $V = e^{\kappa v}$. The Killing horizon is generated by the Killing vector field $k = \kappa \left( V \dd_V - U \dd_U \right)$ and the conventional gauge field (vanishing at infinity) is
\be
\pf{A} = -\frac{Q}{r} \, \df t = -\frac{Q}{2\kappa r}  \left( \frac{1}{V}\,\df V - \frac{1}{U}\,\df U \right) \ .
\ee
However, in this gauge $\pf{A}$ is divergent at the bifurcation surface $(U,V) = (0,0)$. On the other hand, we can choose different gauge,
\be
\pf{A}' = -\frac{Q}{2\kappa} \left( \frac{1}{r} - \frac{1}{r_+} \right)  \left( \frac{1}{V}\,\df V - \frac{1}{U}\,\df U \right) \ ,
\ee
where $r_+$ is the radius of the outer horizon, to obtain regular $\pf{A}$ on the horizon. Likewise, we shall pursue here an alternative gauge choice, in which $\pf{A}$ is finite and smooth at $H[\chi]$ and $\Phi$ vanishes at infinity\footnote{We note in passing that there is also an alternative procedure \cite{Gao03} with a Cauchy surface $\Sigma$ which does not intersects the horizon $H[\xi]$ at the bifurcation surface, but we shall not utilize it here.}. Thus, $i_\xi \pf{A} |_\mathcal{B} = 0$, so that $-i_\xi \pf{A} = \Phi - \Phi_\uH$ (i.e.~$\Phi_0 = \Phi_\uH$) and $i_\xi \pf{A}|_\infty = \Phi_\uH$. With this choice, the $\pf{Q}_\xi^\scr{(em)}$ term drops at the bifurcation surface, but makes contribution at infinity,
\be
\delta\!\oint_{S_\infty} \pf{Q}^\scr{(em)}_\xi = -\Phi_\uH \, \delta Q \ .
\ee
Thus, (\ref{first.inter}) lead to the final form of the first law of black hole mechanics,
\be\label{first}
\delta M = \frac{\kappa}{8\pi} \, \delta \mathcal{A} + \Omega_\uH \, \delta J + \Phi_\uH \delta Q + K_\chi^i \, \delta\beta_i
\ee
with 
\be\label{KLxi}
K_\chi^i \defeq -\frac{1}{4\pi} \int_\Sigma \LL_{\beta_i} \, {\hdg\bm{\chi}} \ .
\ee
An important lesson here is that $K_\chi^i$ appears as a thermodynamic variable conjugate to $\beta_i$. In Sec.~\ref{sec.Smarr} we shall demonstrate that this form of the first law is consistent with the generalized Smarr formula.

\smallskip

The first law obtained in (\ref{first}) does not contain the $\Psi_\uH \, \delta P$ term, which is occasionally included for generality. The formal reason for its absence is that the gauge field $\pf{A}$ is tacitly assumed to be globally well defined and the whole procedure of the covariant phase space formalism should be carefully reexamined to adopt it for solutions with magnetic charge. The only reference, known to us, which has addressed this problem \cite{Keir14}, takes into account contributions on the edges of the local spacetime patches with a well-defined gauge field. These issues are seemingly absent in the approach (1a) to the first law, rendering $\Psi_\uH \, \delta P$ term \cite{Heusler,ZG18}, but it is not clear if any of the aforementioned formal issues have been just swept under the rug. From another point of view \cite{Prabhu15}, the magnetic charge $P$ is a topological charge and it should not vary under perturbations, nor contribute to the first law.

\smallskip

Some of the earlier analyses of the first law of black hole thermodynamics in the presence of NLE fields propose the form of the law with a suspicious absence of the $K^i_\xi \, \delta\beta_i$ term. For example, Herdeiro and Radu \cite{HR19} looked at nonrotating, dyonic black holes in theory with the NLE Lagrangian $\LL = \LL^\scr{(Max)} + \alpha \GG^2$ and derived the first law in the form $\delta M = \kappa\delta\mathcal{A}/(8\pi) + \Phi_\uH \, \delta Q + \Psi_\uH \, \delta P$. However, this result has to be taken with a grain  of salt, as the variation ``$\delta$'' used here keeps the product $\alpha P^2$ fixed. Similarly, one could write the restricted first law $\hat{\delta} M = \kappa\hat{\delta}\mathcal{A}/(8\pi)$ for perturbations with uncharged, nonrotating matter and the corresponding variation $\hat{\delta}$.

\smallskip

Following the recent development of the black hole thermodynamics with the cosmological constant \cite{KRT09,KRT10,KRT11}, one is inclined to interpret the black hole mass $M$ in the first law (\ref{first}) as a generalized ``enthalpy'', related to the internal energy $\mathcal{E}$ via Legendre transformation $M = \mathcal{E} + K_\chi^i \, \beta_i$, so that
\be
\delta \mathcal{E} = \frac{\kappa}{8\pi} \, \delta \mathcal{A} + \Omega_\uH \, \delta J + \Phi_\uH \, \delta Q + \beta_i \, \delta K_\chi^i \ .
\ee
It is not quite clear what is the proper, general interpretation of the quantity $K_\xi^i$. Given that the Lagrangian is written such that the coupling parameter $\beta_i$ has the same physical dimension as $\FF^{1/2}$ (e.g.~$\beta = b$ in Born--Infeld and $\beta = m_e^2/\alpha$ in Euler--Heisenberg theory), that is dimension of the electric field, associated $K_\xi^i$ may be interpreted, based on dimensional argument, as a ``NLE vacuum polarization'' (this was remarked in \cite{GKM12} for the Born--Infeld theory).

\smallskip

Let us now turn to different approach to the first law of black hole mechanics, the physical process version.

\subsection{Physical process first law} 

\begin{figure}
\centering
\begin{tikzpicture}
\draw[thick,lightgray,fill=lightgray,opacity=0.4] (1.5,0.16) [out=90,in=-80] to (1.5,1.5) -- (2.5,2.5) [out=-80,in=90] to (3,0) [out=170,in=0] to (1.5,0.16);
\draw[very thick,gray] (7,3) [out=175,in=5] to (5,3) [out=185,in=-5] to (3,3) -- (0,0) [out=10,in=170] to (3,0) [out=-10,in=190] to (6,0); 
\node at (5.2,-0.4) {\small $\Sigma_0$};
\node at (6,3.3) {\small $\Sigma_1$};
\node at (0.65,1.0) {\small $\mathcal{H}$};
\end{tikzpicture}

\caption{Spacetime diagram of infalling matter, denoted by gray area.} \label{fig:fall}
\end{figure}
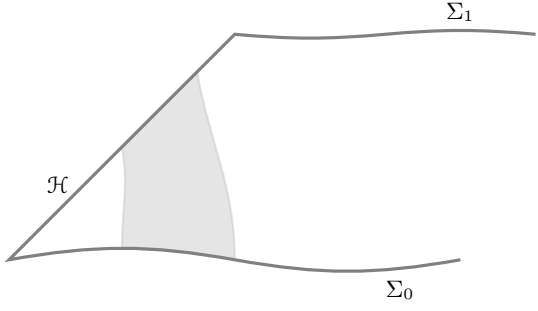

Instead of looking at stationary black hole configurations which are ``nearby'' in some abstract phase space, here we want to describe physical process in which a (possibly charged) matter is thrown into a black hole. Geometric setting is the same as above, except that the Killing horizon $H[\xi]$ no longer needs to be of the bifurcate type. Suppose that $\Sigma_0$ and $\Sigma_1$ are, respectfully, initial and final smooth, spacelike, asymptotically flat Cauchy surfaces, both of which terminate on the horizon $H[\xi]$, as sketched in the Fig.~\ref{fig:fall}. For convenience, the portion of the horizon between $\Sigma_0 \cap H[\xi]$ and $\Sigma_1 \cap H[\xi]$ may be denoted by $\mathcal{H}$. We start from an initial stationary black hole, then perturb it by throwing a small amount of charged matter, and wait until it eventually settles to a final stationary state. Formally, the charged matter is described by fields with compact support which intersects $\Sigma_0$ and $H[\xi]$, but is disjoint from $\Sigma_0 \cap H[\xi]$ (matter is initially away from the black hole) and $\Sigma_1$ (after the process is over and matter has fallen into the black hole, there is no remaining matter on the final Cauchy hypersurface). In addition, we assume that the outward pointing vector field $n^a$ and the corresponding induced orientation $\hat{\veps}$ have been introduced on each of these hypersurfaces, as described in Appendix \ref{app.Stokes}.

Sources are now described by the electromagnetic 4-current $j^a$ and the total energy-momentum tensor $T^\scr{(tot)}_{ab}$, which is a sum of the electromagnetic contribution $T_{ab}$ and the nonelectromagnetic contribution $\widetilde{T}_{ab}$. This generalizes the gNLE field equations to
\be
G_{ab} - 8\pi T_{ab} = 8\pi \widetilde{T}_{ab} \qqd \nabla_{b} Z^{ab} = 4\pi j^{a} \ .
\ee
We assume that $(g_{ab}, \pf{A})$ is a solution of the source-free gNLE equations (\ref{eq:Einst})-(\ref{eq:NLEMax}), while the perturbations $(\delta g_{ab},\delta\pf{A})$ are solutions of the linearized equations with sources $\delta\widetilde{T}_{ab}$ and $\delta j^a$,
\be
\delta (G_{ab} - 8\pi T_{ab}) = 8\pi\delta \widetilde{T}_{ab} \qqd \delta (\nab{b} Z^{ab}) = 4\pi \delta j^a \ .
\ee
Luckily, we do not need to start from scratch, as the expressions for generic variations were prepared within the covariant phase space formalism above. Taking into account that
\begin{align}
\delta (A_a \nab{r} Z^{re}) & = (\delta A_a) \nab{r} Z^{re} + A_a \delta \nab{r} Z^{re} = \nonumber\\
 & = 0 - 4\pi A_a \delta j^e
\end{align}
we see that the variation of the auxiliary 4-form\footnote{The 4-form $\pf{C}$ in \cite{GW01} is written as $\pf{C}_a$, but ``$a$'' is the \emph{last} index, $C_{bcda}$.} $\pf{C}$ (\ref{C}) does not vanish on-shell but
\be
\delta C_{abcd} \approx \left( \delta\tensor{\widetilde{T}}{_a^e} + A_a \delta j^e \right) \epsilon_{ebcd} \ .
\ee
Now, using (\ref{MQSinfty}) and assumption that field perturbations vanish at $\Sigma_0 \cap H[\chi]$, Eq.~(\ref{Omega}) for the Killing vector field $\xi^a = \chi^a$ leads to an on-shell relation
\be\label{eq:deltaMJalpha}
\delta M - \Omega_\uH \, \delta J - K_\chi^i \, \delta\beta_i = - \int_{(\Sigma_0,-\hat{\veps})} {\star\bm{\alpha}}_\chi \ .
\ee
Here we have introduced an auxiliary 1-form $\bm{\alpha}_\xi$, defined by
\be
\bm{\alpha}_\xi \defeq {\hdg(i_\xi\delta\pf{C})} = \delta \widetilde{\pf{T}}(\xi) + (i_\xi\pf{A}) \, \delta\pf{j}
\ee
for any Killing vector field $\xi^a$. Note that the orientation of the hypersurface $\Sigma_0$, emphasized in (\ref{eq:deltaMJalpha}), is opposite of the induced Stokes' orientation $\hat{\veps}$. The symmetry inheritance of all fields and perturbations leads to the conservation of $\bm{\alpha}_\xi$ in a sense that 
\be
\df{\star\bm{\alpha}}_\xi = \df i_\xi \delta\pf{C} = (\Lie_\xi - i_\xi\df) \delta\pf{C} = 0 \ .
\ee
For simplicity, we may suppress the additional index on $\bm{\alpha}$ in what follows. Using the Stokes' theorem (\ref{Stokes.div}) on four-dimensional submanifold bounded by hypersurfaces $\Sigma_0$ and $\Sigma_1$, horizon portion $\mathcal{H}$, and some timelike hypersurface $S$ on which perturbations $\delta \widetilde{T}_{ab}$ and $\delta j_a$ vanish (far away from the black hole), we have
\be
0 = \int_{(\Sigma_0, \hat{\veps})}\!\!(\tilde{n}^a \alpha_a) \hat{\veps} + \int_{(\mathcal{H}, \hat{\veps})}\!\!(-\ell^a \alpha_a) \hat{\veps} \ .
\ee
As we shall deal with the Raychaudhuri equation, a convenient choice for the null vector field $\ell^a$ is $\ell^a = \zeta^a$, a vector field tangent to the \emph{affinely} parametrized null generators of the unperturbed Killing horizon $H[\xi]$. Taking into account all these remarks, we may ``shift'' the integral in (\ref{eq:deltaMJalpha}) from $\Sigma_0$ to the black hole horizon,
\begin{align}
-\int_{(\Sigma_0,-i_n \veps)}\!\!\!\!{\hdg\bm{\alpha}} & = -\int_{(\Sigma_0,-i_n \veps)}\!\!\!\!(-n^a \alpha_a)(i_n\veps) = \nonumber\\
 & = \int_{(\mathcal{H},i_n\veps)} (\zeta^a \alpha_a)(i_n \veps) \ ,
\end{align}
where we have, for simplicity, left out the pullback symbols. In other words, with assumed induced orientation of the horizon, we have
\be
\delta M - \Omega_\uH \, \delta J - K_\chi^i \, \delta\beta_i = \int_\mathcal{H} \zeta^a \alpha_a \, \hat{\veps} \ .
\ee
There are two contributions to this integral, electromagnetic and nonelectromagnetic. For the evaluation of the former one, we use the gauge choice in which both $\Phi$ and $\pf{A}$ are zero at infinity, so that $\Phi_0 = 0$ and $-i_\xi\pf{A} = \Phi_\uH$ on the horizon. For the causal, future directed $\delta j^a$, we have $\zeta^a \delta j_a \le 0$ on the horizon, corresponding to the positive amount of the infalling charge, $\delta Q \ge 0$ (and vice versa for the negatively charged infalling matter). This gives us 
\be
\delta M - \Omega_\uH \, \delta J - \Phi_\uH \, \delta Q - K_\chi^i \, \delta\beta_i = \int_\mathcal{H} \zeta^a \chi^b (\delta \widetilde{T}_{ab}) \, \hat{\veps} \ .
\ee
It remains to be shown that the right-hand side is the area term in the first law.

\smallskip

The Raychaudhuri equation for the null congruence\footnote{For the extremal Killing horizon $H[\chi]$, with $\kappa = 0$, the Killing vector field $\chi^a$ is automatically tangent to the affinely parametrized null geodesic generators of the horizon, thus $\zeta^a = \chi^a$.} $\zeta^a = \chi^a/(\kappa V)$, with the corresponding affine parameter $V$, 
\be
\frac{\df\theta}{\df V} = -\frac{1}{2}\,\theta^2 - \sigma_{ab}\sigma^{ab} - R_{ab}\zeta^{a}\zeta^{b} \ ,
\ee
in combination with vanishing of the expansion ($\theta = 0$) and shear ($\sigma_{ab} = 0$) for the stationary background, and Einstein field equation, reduces to
\be 
\frac{\df\theta}{\df V} = -8\pi \left( T_{ab} + \widetilde{T}_{ab} \right) \zeta^a \zeta^b \ .
\ee
In order to extract the change in area of the black hole horizon, we need to look at the perturbed Raychaudhuri equation. Diffeomorphism freedom allows us to make the gauge choice such that null generators of the unperturbed and perturbed black hole horizons coincide, which amounts to $\delta\zeta^a \sim \zeta^a$ on the horizon. Thus, using the fact \cite{Wald} that $R_{ab} \zeta^a \zeta^b |_H = 0$, the perturbed Raychaudhuri equation \cite{GW01} is
\be\label{eq:pertRay}
\frac{\df\delta\theta}{\df V} = -8\pi \left( \delta T_{ab} + \delta\widetilde{T}_{ab} \right) \zeta^{a}\zeta^{b} \big|_H \ .
\ee
The first term on the right-hand side consists of
\begin{align}
\delta T_{ab} \zeta^a \zeta^b & = -4 (\delta\LL_\FF) T_{ab}^{\scr{(Max)}} \zeta^a \zeta^b - \nonumber\\
 & - 4\LL_\FF \, \delta T_{ab}^{\scr{(Max)}} \zeta^a \zeta^b + \frac{1}{4} \, \delta (T g_{ab}) \zeta^a \zeta^b \ .
\end{align}
Using the fact that $\zeta^a$ is null both in the unperturbed and perturbed spacetimes, the last term is immediately zero on the horizon, while
\begin{align}
4\pi T_{ab}^{\scr{(Max)}} \zeta^a \zeta^b |_H & = (\kappa V)^{-2} E_a E^a |_H = 0 \\
4\pi \delta T_{ab}^{\scr{(Max)}} \zeta^a \zeta^b |_H & = (\kappa V)^{-2} \delta (E^a E_a) |_H = 0
\end{align}
due to fact, established for the zeroth law, that the electric field $E^a$ is null on the horizon. The remaining perturbed Raychaudhuri equation (\ref{eq:pertRay}) may be put in the following form
\be
\kappa V \frac{\df\delta\theta}{\df V} = -8\pi \zeta^a \chi^b \delta \widetilde{T}_{ab} \big|_H \ .
\ee
Integral of the left-hand side along the horizon portion $\mathcal{H}$ leads \cite{WaldQFT} to the change in area $\delta\mathcal{A}$,
\be
\int_\mathcal{H} \zeta^a \chi^b (\delta \widetilde{T}_{ab}) \, \hat{\veps} = \frac{\kappa}{8\pi} \, \delta\mathcal{A} \ .
\ee
Putting all this together, we have obtained the physical process first law if the black hole mechanics,
\be
\delta M = \frac{\kappa}{8\pi} \, \delta \mathcal{A} + \Omega_\mathsf{H} \, \delta J + \Phi_\mathsf{H} \, \delta Q + K_\chi^i \, \delta\beta_i \ ,
\ee
consistent with (\ref{first}).

\section{Remarks on the generalized Smarr formula} \label{sec.Smarr} 

The problem of generalization of the Smarr formula for rotating (stationary axially symmetric) black holes in the $\FF\GG$-class NLE theories has been recently solved \cite{GS17}, with an interim result of the form
\be
M = \frac{\kappa}{4\pi}\,\mathcal{A} + 2\Omega_\uH J + \Phi_\uH Q_\uH + \Psi_\uH P_\uH + \frac{1}{2} \int_\Sigma T \, {\hdg\bm{\chi}} \ .
\ee
This relation follows directly from the Bardeen--Carter--Hawking mass formula and is in principle independent of the first law. The additional last term on the right-hand side is clearly absent in the Maxwell's electrodynamics, for which $T = 0$, but does not vanish in general NLE theory. Furthermore, as was observed in \cite{GS17}, if the NLE Lagrangian is of the form $\LL = \sigma^{-1} f(\sigma\FF,\sigma\GG)$, with some parameter $\sigma$ and a real function $f$, then the trace of the energy-momentum tensor may be written as $T = (-\sigma/\pi) \, \dd_\sigma\LL$, allowing us to write the additional NLE term, at least formally, as a product of a conjugate pair of thermodynamic variables.

\smallskip

Nevertheless, once the first law is obtained, we may deduce the Smarr formula using a particular choice of perturbation, that is a path through the phase space of solutions defined by the carefully chosen scaling of fields \cite{SW93}. This approach has been used by Zhang and Gao \cite{ZG18} for the $\FF$-class NLE theories, along a bit of meandering procedure as they derive the first law by variation of the mass formula, approach (1a) mentioned in Sec.~\ref{sec:first}. We shall rederive the Smarr formula from the first law (\ref{first}) in order to check the consistency of the complete procedure.

\subsection{Smarr formula from the first law} 

Let $(g_{ab}, \pf{A})$ be an initial solution of the gNLE field equations. Our first aim is to find a family of rescaled field configurations $(\lambda^2 g_{ab}, \lambda^\nu \pf{A})$ with a real parameter $\lambda$ and a real constant $\nu$ chosen such that rescaled fields are again solutions. Of course, there is no \emph{a priori} guarantee that such simple construction is possible, but we shall prove that this is indeed the case. Also, note that the Smarr formula is sometimes obtained via Euler's theorem for homogeneous functions \cite{Lee}, under the assumption that the black hole mass $M(\mathcal{A},J,Q,\dots)$ is a differentiable homogeneous function of degree $1$. Eulerian approach was, in fact, used in the original Smarr's derivation \cite{Smarr72} for the Kerr--Newman black hole. However, any generalization in this approach demands a careful justification of the homogeneity of the black hole mass function for a theory under consideration, as it does not hold in general \cite{HKO19}.

\smallskip

Let us now carefully examine the scaling of all objects appearing in our analysis of the spacetime. Metric rescaling $g_{ab} \to \lambda^2 g_{ab}$ immediately implies corresponding rescaling for the inverse metric, $g^{ab} \to \lambda^{-2} g^{ab}$, volume form, $\veps \to \lambda^4 \veps$, area of the black hole horizon, $\mathcal{A} \to \lambda^2 \mathcal{A}$, as well as the Riemann tensor and its contractions,
\be
\tensor{R}{^a_b_c_d} \to \tensor{R}{^a_b_c_d} \,,\ R_{ab} \to R_{ab} \,,\ R \to \lambda^{-2} R \,,\ G_{ab} \to G_{ab} \nonumber
\ee
Killing vector $k^a$ is normalized at infinity via $g_{ab} k^a k^b = -1$, so that $k^a \to \lambda^{-1} k^a$ and $\pf{k} \to \lambda \pf{k}$. Killing vector $m^a$ is normalized along its closed orbits $\mathcal{C}$,
\be
\oint_\mathcal{C} \frac{1}{m_a m^a} \, \pf{m} = 2\pi
\ee
so that $m^a \to m^a$ and $\pf{m} \to \lambda^2 \pf{m}$. In order to have consistent combination $\chi^a = k^a + \Omega_\uH m^a$, we need $\Omega_\uH \to \lambda^{-1} \Omega_\uH$. Also, using the geodesic equation for the Killing vector field $\xi^a$ generating the Killing horizon, $\xi^b \nab{b} \xi^a = \kappa \xi^a$, we have $\kappa \to \lambda^{-1} \kappa$ for the surface gravity $\kappa$. Consequently, via Komar integrals (\ref{KomarMJ}), we know that $M \to \lambda M$ and $J \to \lambda^2 J$.

\smallskip

Now we turn to the gauge sector. Given that the gauge field scales as $\pf{A} \to \lambda^\nu \pf{A}$ and using the metric scaling described above, we immediately have $\pf{F} \to \lambda^\nu \pf{F}$, ${\hdg\pf{F}} \to \lambda^\nu {\hdg\pf{F}}$, as well as $\FF \to \lambda^{2(\nu-2)} \FF$ and $\GG \to \lambda^{2(\nu-2)} \GG$. For the electric and magnetic 1-forms defined with respect to the Killing vector field $\chi^a$, we have $\pf{E} \to \lambda^{\nu - 1} \pf{E}$ and $\pf{B} \to \lambda^{\nu - 1} \pf{B}$, so that the associated scalar potentials scale as $\Phi \to \lambda^{\nu - 1} \Phi$ and $\Psi \to \lambda^{\nu - 1} \Psi$. Einstein's field equation $G_{ab} = 8\pi T_{ab}$ implies that the energy-momentum tensor should be scale invariant, $T_{ab} \to T_{ab}$, and from (\ref{TZL}), we see that one consistent choice is $\LL \to \lambda^{-2} \LL$. By demanding that this scaling is universal, that is valid for all electromagnetic Lagrangians, basic Maxwell's case implies $\nu = 1$. This choice tacitly implies that coupling parameters in a NLE Lagrangian will have some specific scaling, say $\beta_i \to \lambda^{b_i} \beta_i$ for some real exponents $b_i$. For example, we have $b \to \lambda^{-1} b$ in the Born--Infeld theory and $\alpha \to \lambda \alpha$ in the Euler--Heisenberg theory. Consequently, from (\ref{KomarQP}), we have $Q \to \lambda Q$ and $P \to \lambda P$, while (\ref{KLxi}) leads to $K_i \to \lambda^{1- b_i} K_i$. All definite scaling exponents deduced above are collected in Table \ref{tab:scale}. We stress that these are not some \emph{necessary} scaling transformations, rather a consistent (and convenient) choice which allows us to apply the first law of black hole mechanics.

\begin{table}
\centering
\begin{tabular}{rl}
Scaling exponent & \\
\hline
$-2$ & $g^{ab}$, $R$, $\FF$, $\GG$ \\
$-1$ & $\kappa$, $\Omega_\uH$ \\
$0$ & $\tensor{R}{^a_b_c_d}$, $R_{ab}$, $G_{ab}$, $\pf{E}$, $\pf{B}$, $\Phi$, $\Psi$ \\
$1$ & $M$, $\pf{k}$, $\pf{A}$, $\pf{F}$, ${\hdg\pf{F}}$, $Q$, $P$ \\
$2$ & $g_{ab}$, $\pf{m}$, $\mathcal{A}$, $J$ \\
$4$ & $\veps$
\end{tabular}
\caption{A summary of scaling exponents for various fields and charges.} \label{tab:scale}
\end{table}

\smallskip

All quantities varied in the first law of black hole mechanics are functions of the parameter $\lambda$ of the form
\be
\mathscr{Q}(\lambda) = \lambda^q \mathscr{Q}(1) \ ,
\ee
with some scaling exponent $q$. Hence, we have
\be
\delta\mathscr{Q} = \frac{\df \mathscr{Q}(\lambda)}{\df\lambda} \Big|_{\lambda=1} = q\mathscr{Q} \ ,
\ee
where we have used abbreviation $\mathscr{Q} = \mathscr{Q}(1)$ for the initial, unperturbed quantity. Putting all this together we can recover the generalized Smarr formula
\be
M = \frac{\kappa}{4\pi} \, \mathcal{A} + 2\Omega_\uH J + \Phi_\uH Q + \sum_i b_i K^i_\chi \beta_i \ .
\ee
Again, as was remarked under Eq.~(\ref{first}), the absence of the $\Psi_\uH P$ term in this procedure is a consequence of its absence in our form of the first law. On the other hand, direct derivation of the generalized Smarr formula \cite{GS17}, being independent of the first law, evades these obstacles and contains  the magnetic potential-charge term.

\smallskip

The authors in \cite{ZG18} argue that the Smarr formula obtained via scaling argument is of greater generality since it may treat the NLE Lagrangians with multiple coupling parameters. However, the only such example known to us is the ABG Lagrangian (\ref{ABG}) and even here, as was already remarked in \cite{GS17}, one might write the Lagrangian in a form $\LL = \tilde{\mu} \alpha^{-1} f(\alpha\FF)$, with $\tilde{\mu} = \mu/g$ and $\alpha = g^2$. Then, as the parameters scale as $\mu \to \lambda \mu$ and $g \to \lambda g$, the parameter $\tilde{\mu}$ is scale invariant, implying that ABG case is still covered by the procedure presented in \cite{GS17}. Even more generally, one might argue that any physically sensible NLE theory should in some weak field limit be of the form
\be
\LL = -\frac{1}{4} \, \FF + \sigma \left( c_{20} \FF^2 + 2c_{11} \FF\GG + c_{02} \GG^2 \right) + O(\sigma^2)
\ee
with dominant Maxwell's term and expansion in some coupling parameter $\sigma$ (dimensionless constants $c_{ij}$ are irrelevant here). Then, a simple algebraic manipulation,
\begin{align}
\LL & = \frac{1}{\sigma} \Big( -\frac{1}{4} \, (\sigma\FF) + c_{20} (\sigma\FF)^2 + \nonumber\\
 & + 2c_{11} (\sigma\FF)(\sigma\GG) + c_{02} (\sigma\GG)^2 + O(\sigma^3) \Big)
\end{align}
brings such Lagrangian in a form which was discussed in \cite{GS17}. Note that in this case the scaling of the coupling parameter is $\sigma \to \lambda^2 \sigma$.

\subsection{Linearity of the Smarr formula} 

Finally, we turn to the question about the (non-)linearity of the Smarr formula: for which NLE theories the generalized Smarr formula may be brought to the form
\begin{align}
c_1 M = c_2\kappa\mathcal{A} & + c_3\Omega_\uH J + c_4\Phi_\uH Q + \nonumber\\
 & + c_5\Psi_\uH P + c_6\Phi_\uH P + c_7\Psi_\uH Q
\end{align}
with some real constants $\{c_1,\dots,c_7\}$? A systematic approach to the problem is to look into terms which would, upon integration of the 3-form $T\,{\hdg\bm{\chi}}$ over $\Sigma$, produce such products of potentials and charges,
\begin{align}
\df (\Phi \, {\hdg\pf{Z}}) & = -\pf{E} \w {\hdg\pf{Z}} = \frac{1}{2}\,{\hdg\pf{R}}(\chi) + \left( 2\pi T - \LL \right) {\hdg\bm{\chi}} \\
\df (\Psi \pf{F}) & = -\pf{H} \w \pf{F} = \frac{1}{2}\,{\hdg\pf{R}}(\chi) + \LL\,{\hdg\bm{\chi}} \\
\df (\Phi \pf{F}) & = \frac{1}{2}\,i_\chi (\pf{F} \w \pf{F}) = -\frac{1}{4}\,\GG\,{\hdg\bm{\chi}} \\
\df (\Psi \, {\hdg\pf{Z}}) & = -\frac{1}{2}\,i_\chi ({\hdg\pf{Z}} \w {\hdg\pf{Z}}) = \nonumber\\
 & = 4 \left( 2\LL_\FF\LL_\GG \FF + (\LL_\GG^2 - \LL_\FF^2) \GG \right) {\hdg\bm{\chi}}
\end{align}
These equations deserve a brief explanation. The first is obtained from the Einstein field equation, energy-momentum tensor in the form (\ref{TZL}) and identity ${\hdg i_E \pf{Z}} = -\pf{E} \w {\hdg\pf{Z}}$. The second is obtained by combining the first one with contraction of (\ref{eq:FwhZ}) with the Killing vector field $\chi^a$. The remaining two equations are obtained by contractions of (\ref{eq:FwF}) and (\ref{eq:hZwhZ}) with $\chi^a$.

Upon inspection, it is suggestive, although we do not have a watertight argument, that a necessary condition for the linearity of the Smarr formula is
\be\label{linSmarrL}
\LL = a( \LL_\FF \FF + \LL_\GG \GG ) + b \left( 2\LL_\FF\LL_\GG \FF + (\LL_\GG^2 - \LL_\FF^2) \GG \right) + c\GG
\ee
for some real constants $a$, $b$, and $c$. Namely, this allows one to turn a linear combination of ${\hdg\pf{R}(\chi)}$ and $T\,{\hdg\bm{\chi}}$ into a linear combination of terms $\df(\Phi\,{\hdg\pf{Z}})$, $\df(\Psi \pf{F})$, $\df(\Phi\pf{F})$, and $\df(\Psi\,{\hdg\pf{Z}})$, with cancellation of the remaining terms. As the term $\GG$ is nondynamical, we can dispose of it and set $c=0$. The remaining condition may be interpreted as a nonlinear partial differential equation for the Lagrangian (as a function of two variables, $\FF$ and $\GG$), but unfortunately we do not know its complete, general solution.

\smallskip

One possible simplification may be obtained if we restrict the analysis to NLE theories which are invariant with respect to electromagnetic duality rotation, defined by $\pf{F} \to \pf{F} \cos\alpha + {\hdg\pf{Z}} \sin\alpha$ and $\pf{Z} \to \pf{Z} \cos\alpha + {\hdg\pf{F}} \sin\alpha$ with a real angle $\alpha$. It is known \cite{GR95} that a necessary and sufficient condition for such invariance to hold is that difference ${\hdg Z_{ab}} Z^{ab} - \GG$ be constant for any field configuration, which translates into constancy of combination $2\LL_\FF\LL_\GG \FF + (\LL_\GG^2 - \LL_\FF^2) \GG + (\GG/16)$. This, in turn, implies that the linearity of the Smarr formula in any duality invariant NLE theory simplifies to the linear, $b = 0 = c$ case. Characteristics of the partial differential equation $\LL = a( \LL_\FF \FF + \LL_\GG \GG )$ in the $\FF$-$\GG$ plane, defined by the system $(\dot{\FF},\dot{\GG}) = (\FF,\GG)$, are nothing but lines through the origin. The partial differential equation is reduced, along a characteristic, to the ordinary differential equation $a\dot{\LL} - \LL = 0$. Hence, on a domain where $\FF \ne 0$, we can write the general solution in a form $\LL = \FF^{1/a} \, f(\GG/\FF)$, while on a domain where $\GG \ne 0$ in a form $\LL = \GG^{1/a} \, g(\FF/\GG)$, with some differentiable real functions $f$ and $g$. A prominent class of examples are all NLE theories with traceless energy-momentum tensor, solutions of the $(a,b,c) = (1,0,0)$ case, a member of which is recently introduced ModMax theory \cite{BLST20,FAGMLM20,BBKP20}. Also, for constant $f$ and $a = 1/s$, we have the power-Maxwell class of NLE theories (linearity of the corresponding Smarr formula has been already confirmed in \cite{GS17}).

\smallskip

Another pragmatic approach is to insist that the NLE Lagrangian should behave close to the Maxwell's in a weak field limit. More precisely, let us assume that Lagrangian $\LL$ is defined on an open subset $O \subseteq \rr^2$, such that (a) $(0,0) \in O$, (b) $\LL : O \to \rr$ is a $C^2$ function, and (c) $\LL_\FF(0,0) = -1/4$ and $\LL_\GG(0,0) = 0$. Then partial derivatives of (\ref{linSmarrL}) with respect to $\FF$ and $\GG$ imply
\be
-\frac{1}{4} = \LL_\FF(0,0) = -\frac{1}{4} \, a \quad \textrm{and} \quad 0 = \LL_\GG(0,0) = -\frac{1}{16} \, b \ , \nonumber
\ee
so that $(a,b) = (1,0)$, leading us back to the linear case of the partial differential equation (\ref{linSmarrL}). Furthermore, let $V \subseteq O$ be an open set star-shaped with respect to the origin (for all $x \in V$ the line segment from the origin to $x$ is contained in $V$), in which we analyse problem along the lines defined by $\GG = p\FF$, with a real parameter $p$. If the solution is written in a form $\LL = \FF f(\GG/\FF)$, then along these lines we have $\LL_\FF = f(p) - pf'(p)$ and $\LL_\GG = f'(p)$, while conditions (b) and (c) above imply that $f(p) = -1/4$ for any $p \in \rr$. Have we used the other form of the solution, $\LL = \GG \, g(\FF/\GG)$, and lines defined by $\FF = p\GG$, analogous reasoning would lead us to the equivalent conclusion that $g(p) = -p/4$ for any $p \in \rr$. In other words, given that (\ref{linSmarrL}) is indeed a necessary condition for the linearity of the Smarr formula (which yet has to be proven rigorously), the only NLE theory with the Maxwellian weak field limit and linear Smarr formula is the Maxwell's electrodynamics itself, at least on some neighbourhood of the origin of $\FF$-$\GG$ plane.

\section{Discussion} 

The elaborate web of connections between gravitational theories and thermodynamics needs to be tested against all physically motivated modifications of the classical Einstein--Maxwell theory, hoping that this might lead us to some novel insights about the microscopic picture of the spacetime. The main goal of this paper was to complete the basic architecture of the black hole thermodynamics in the presence of NLE fields, above all the first law of black hole mechanics, along with all the auxiliary results that such relation rests upon.

\smallskip

To this end, building on some earlier hints and ideas \cite{MO11a,MO11b,GKM12,HJPY17,GS17,ZG18}, we have extended the covariant phase space approach to the first law of black hole mechanics, both the equilibrium and the physical process versions, in spacetimes with NLE fields. Just as the cosmological constant enters the black hole thermodynamic relations in pair with the conjugate volume, the major novelty here is the introduction of conjugate pairs $(\beta_i,K^i_\xi)$ of NLE Lagrangian parameters $\beta_i$ and ``NLE vacuum polarization'' $K^i_\xi$ among the thermodynamic variables. Also, we have completed several versions of the proof of the zeroth law of black hole electrodynamics, constancy of the scalar potentials on the horizon, which is an essential ingredient for the other laws. This has allowed us to generalize the first law for rotating black holes in $\FF\GG$-class of NLE theories, which can now be applied to theories with QED corrections of the Maxwell's electrodynamic Lagrangian. Furthermore, in order to prove the consistency of results obtained here with the NLE Smarr formula \cite{GS17}, we have derived the Smarr formula from the first law, using the so-called scaling approach. Finally, we have presented an argument that the linear form of the Smarr formula in $\FF\GG$-class of NLE theories appears only in Maxwell's theory or NLE theories which do not possess a Maxwellian weak field limit. 

\smallskip

Some of the generalizations that wait ahead are pretty much straightforward. For example, inclusion of the cosmological constant, with the additional $V \df\Lambda$ term in the first law, can be achieved according to an established procedure \cite{KRT09,KRT10,KRT11,KMT17}. Extensions of the first law for gravitational theories beyond the canonical general relativity, as long as the electromagnetic field is minimally coupled to gravitation, are in principle covered by the covariant phase space formalism procedures \cite{Wald93,IW94,Tachikawa06,BCDPPS11}, although a concrete evaluation of the corrections may be a formidable task. Nontrivial contributions to the gravitational Einstein--Hilbert action can appear, for instance, due to quantization (in a sense of an effective theory) \cite{Donoghue94,BOS,CCK20} or quantum gravity (via spectral triple) \cite{CC96,MPR20}. One line of future developments are generalizations for the lower and higher dimensional spacetimes, with caveat that invariant $\GG$ has to be excluded or replaced with something else, as $\pf{F}$ and its Hodge dual ${\hdg\pf{F}}$ have equal orders only in four spacetime dimensions. Considerably bigger challenge is to generalize all these results for NLE Lagrangians which also include terms with covariant derivatives of the 2-form $\pf{F}$, as well as nonminimal coupling to gravitation and other matter fields. Such corrections to the Maxwell's Lagrangian could be produced via generalized uncertainty principle \cite{BDT20} or induced from the noncommutative field theories \cite{MSSW00,HJM11,CDKS18,GJSS19}.

\smallskip

There is yet another intriguing relation which should be fully resolved and better understood. Namely, it has been recently observed \cite{AOORG18,DOO19} that field redefinitions admit establishment of mapping (``frame change'') between the (a) so-called Eddington-inspired Born--Infeld gravitational theory \cite{BZ10} 
\be
\LL^\scr{(EiBI)} = \frac{1}{\kappa^2\varepsilon} \, \sqrt{|\det(g_{\mu\nu} + \varepsilon R_{\mu\nu})|}
\ee
coupled to Maxwell's electromagnetism, and (b) Einstein--Hilbert gravitational theory coupled to Born--Infeld nonlinear electromagnetism, with Lagrangian $\LL^\scr{(BI)}$. The question is whether it is possible to implement this correspondence directly to the first law of black hole thermodynamics, that is, can we relate $\dd\LL^\scr{(EiBI)}/\dd R_{abcd}$ in the Wald's entropy formula \cite{Wald93,IW94} and $\dd\LL^\scr{(BI)}/\dd b$ in the NLE term $b\,{\delta K}$, given that the former appears in an integral over a 2-surface, while the latter is part of the integral over the hypersurface. A hope that such relation is feasible comes from the fact that field redefinition \cite{DOO19} comprises parameter correspondence of the form $b^2 = -1/(2\varepsilon\kappa^2)$. We leave this inquiry for the future work.

\smallskip

It remains to be seen if extension of the phase space and additional variations of the Lagrangian parameters in the first law are a mere algebraic, bookkeeping device, or an important hint for understanding of the thermodynamic features of the spacetime.

\appendix
\section{Important identities} 

Here we collect several basic definitions and identities with differential forms, used throughout the paper. Suppose that $\bm{\omega}$ is a $p$-form on a smooth Lorentzian four-dimensional manifold. Then the Hodge dual, contraction with vector $X^a$ and exterior derivative $\df$ are, respectively, defined as
\begin{align}
(\hdg\omega)_{a_{p+1} \dots a_4} & = \frac{1}{p!}\,\omega_{a_1 \dots a_p} \tensor{\epsilon}{^{a_1}^{\dots}^{a_p}_{a_{p+1}}_{\dots}_{a_4}} \\
(i_X \omega)_{a_1 \dots a_{p-1}} & = X^b \omega_{b a_1 \dots a_{p-1}} \\
(\df\omega)_{a_1 \dots a_{p+1}} & = (p+1) \nab{[a_1} \omega_{a_2 \dots a_{p+1}]}
\end{align}
Hodge dual twice applied is identity up to sign, ${\hdg{\hdg\bm{\omega}}} = (-1)^{p(4-p) + 1} \bm{\omega}$ (plus for odd $p$ and minus for even $p$). We immediately have ${\hdg 1} = \veps$ and ${\hdg\veps} = -1$. Particularly useful operation is the so-called ``flipping over the Hodge'',
\be
i_X {\hdg\bm{\omega}} = {\hdg(\bm{\omega} \w \pf{X})}
\ee
where $\pf{X}$ is the associated 1-form, $X_a = g_{ab} X^b$. Note that $i_X \veps = {\hdg\pf{X}}$. For any 1-form $\bm{\alpha}$, we have ${\hdg\df{\hdg\bm{\alpha}}} = -\nabla^a \alpha_a$ and $\df {\hdg\bm{\alpha}} = (\nabla^a \alpha_a) \veps$. 

\smallskip

For any 2-form $\pf{F}$, we have two essential identities
\begin{align}
F_{ac} \tensor{F}{^c_b} - {\hdg F}_{ac} \tensor{{\hdg F}}{^c_b} & = -\frac{1}{2} \, \FF g_{ab} \ , \label{eq:FFHFHF} \\
F_{ac} \, \tensor{{\hdg F}}{^c_b} = {\hdg F}_{ac} \, \tensor{F}{^c_b} & = -\frac{1}{4} \, \GG g_{ab} \ . \label{eq:FHF}
\end{align}
Furthermore, using the identity $2 \bm{\alpha} \w {\hdg\bm{\beta}} = (\alpha_{ab}\beta^{ab})\veps$, valid for any 2-forms $\bm{\alpha}$ and $\bm{\beta}$, it is straightforward to derive the following identities
\begin{align}
\pf{F} \w {\hdg\pf{F}} & = \frac{1}{2}\,\FF \, \veps \label{eq:FwhF} \\
\pf{F} \w \pf{F} & = -\frac{1}{2}\,\GG \, \veps \label{eq:FwF} \\
\pf{F} \w {\hdg\pf{Z}} & = -2 (\FF\LL_\FF + \GG\LL_\GG) \, \veps \label{eq:FwhZ} \\
\pf{F} \w \pf{Z} & = -2 (\FF\LL_\GG - \GG\LL_\FF) \, \veps \label{eq:FwZ} \\
{\hdg\pf{Z}} \w {\hdg\pf{Z}} & = 8 \left( (\LL_\FF^2 - \LL_\GG^2) \GG - 2\LL_\FF\LL_\GG \FF \right) \veps \label{eq:hZwhZ}
\end{align}
Finally, taking into account ${\hdg F}^{ab} \nab{c} F_{ab} = F^{ab} \nab{c} \, {\hdg F}_{ab}$ and assuming that $\df\pf{F} = 0$, we have
\be
F^{ac} \nab{a} F_{bc} = \frac{1}{4} \, \nab{b} \FF \label{eq:FnabF}
\ee
and
\be
{\hdg F}^{ac} \nab{a} F_{bc} = F^{ac} \nab{a} {\hdg F}_{bc} = \frac{1}{4} \, \nab{b} \GG \ . \label{eq:HFnabF}
\ee

\section{Stokes' theorem on Lorentzian manifolds} \label{app.Stokes} 

Suppose that $\mathscr{M}$ is an orientable smooth $m$-manifold with boundary $\dd\mathscr{M}$ and inclusion $\imath : \dd\mathscr{M} \hookrightarrow \mathscr{M}$. An orientation on $\mathscr{M}$ is fixed by choice of a nowhere vanishing $m$-form $\veps$, while corresponding induced (Stokes) orientation on the boundary is defined as $\hat{\veps} = \imath^*(i_N \veps)$, with the \emph{outward pointing} nonvanishing vector field $N^a$ on $\dd\mathscr{M}$. For any smooth, compactly supported $(m-1)$-form $\bm{\alpha}$ on $\mathscr{M}$, the Stokes' theorem \cite{Lee} states that
\be\label{Stokes}
\int_{(\mathscr{M},\veps)}\!\!\df\bm{\alpha} = \int_{(\dd\mathscr{M},\hat{\veps})}\!\!\imath^*\bm{\alpha} \ ,
\ee
where we have, for clarity, emphasized orientations for both the manifold and its boundary. Although the Stokes' theorem does not rely on any additional structure on the manifold, such as metric or connection, it admits some practical corollaries on (pseudo)-Riemannian manifolds. Suppose that $\mathscr{M}$ is a smooth Lorentzian manifold and $\mathscr{N} \subseteq \mathscr{M}$ its embedded compact $m$-dimensional submanifold with boundary $\dd\mathscr{N}$, inclusion $\jmath : \dd\mathscr{N} \hookrightarrow \mathscr{N}$ and an outward pointing, nonvanishing vector field $n^a$ on $\dd\mathscr{N}$. The corresponding induced orientation on the boundary $\dd\mathscr{N}$ is $\hat{\veps} = \jmath^*(i_n\veps)$. Then for any smooth vector field $v^a$ on $\mathscr{N}$, the Stokes' theorem implies
\be
\int_{(\mathscr{N},\veps)}\!\!(\nab{a} v^a) \, \veps = \int_{(\mathscr{N},\veps)}\!\!\df i_v \veps = \int_{(\dd\mathscr{N},\hat{\veps})}\!\!\jmath^* (i_v \veps) \ .
\ee

\medskip

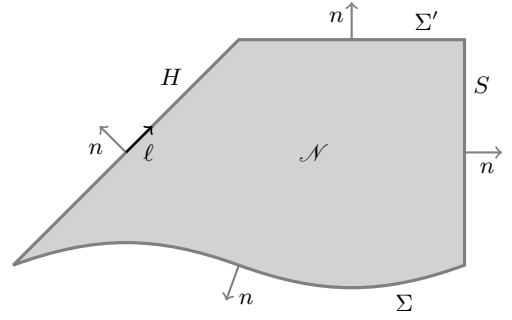
\begin{figure}
\centering
\begin{tikzpicture}
\draw[fill=lightgray,opacity=0.7] (0,0) [out=20,in=160] to (3,0) [out=-20,in=200] to (6,0) -- (6,3) -- (3,3) -- (0,0);
\draw[very thick,gray] (0,0) [out=20,in=160] to (3,0) [out=-20,in=200] to (6,0) -- (6,3) -- (3,3) -- (0,0); 
\draw[gray,thick,->] (6,1.5) -- (6.5,1.5);
\draw[gray,thick,->] (4.5,3) -- (4.5,3.5);
\draw[gray,thick,->] (1.5,1.5) -- (1.15,1.85); 
\draw[thick,->] (1.5,1.5) -- (1.85,1.85); 
\draw[gray,thick,->] (3,0) -- (2.83,-0.47);
\node at (4,1.5) {$\mathscr{N}$};
\node at (5.2,-0.5) {\small $\Sigma$};
\node at (5.5,3.25) {\small $\Sigma'$};
\node at (2.1,2.5) {\small $H$};
\node at (6.23,2.4) {\small $S$};
\node at (1.8,1.5) {\small $\ell$};
\node at (1.1,1.55) {\small $n$};
\node at (3.1,-0.45) {\small $n$};
\node at (4.3,3.3) {\small $n$};
\node at (6.3,1.3) {\small $n$};
\end{tikzpicture}

\caption{Submanifold $\mathscr{N}$ with four parts of the boundary (spacelike hypersurfaces $\Sigma$ and $\Sigma'$, timelike hypersurface $S$, null hypersurface $H$) and corresponding outward pointing vector field $n^a$.} \label{fig:N}
\end{figure}

\noindent
Let us, for concreteness, assume that the boundary of $\mathscr{N}$ consists of two spacelike hypersurfaces $\Sigma$ and $\Sigma'$, timelike hypersurface $S$, and a null hypersurface (portion of a black hole horizon) $H$,
\be
\dd\mathscr{N} = \Sigma \cup \Sigma' \cup S \cup H \ , \nonumber
\ee
as sketched in Fig.~\ref{fig:N}. For each part of the boundary, it is convenient to have a corresponding decomposition of the volume form $\veps$:

\begin{itemize}
\item[(i)] The non-null part of the boundary. We assume that $n^a$ is normalized such that $n^a n_a = \pm 1$. Also, following the convection in \cite{Wald}, we introduce the auxiliary vector field $\tilde{n}^a \defeq (n^b n_b) \, n^a$, so that $\tilde{n}^a$ is outward pointing for spacelike $n^a$ and inward pointing for timelike $n^a$. Then $\pf{n} \w i_n\veps = f \veps$ for some function $f$ and contraction with $n^a$ leads to the decomposition
\be
\veps = (n^a n_a) \, \pf{n} \w i_n\veps = \tilde{\pf{n}} \w i_n\veps \ .
\ee

\item[(ii)] The null part of the boundary generated by the future directed vector field $\ell^a$. For the outward pointing vector field, we take a future directed null vector field $n^a$ on $H$, normalized such that $n^a \ell_a = -1$. Then $\bm{\ell} \w i_n\veps = f \veps$ for some function $f$ and contraction with $n^a$ leads to the decomposition
\be
\veps = -\bm{\ell} \w i_n\veps \ .
\ee
\end{itemize}
 
These decompositions imply
\be
\jmath^* (i_v \veps) = \left\{ \begin{array}{rl} (\tilde{n}_a v^a) \hat{\veps} & \ \textrm{on non-null part of} \ \dd\mathscr{N} \\ -(\ell_a v^a) \hat{\veps} & \ \textrm{on null part of} \ \dd\mathscr{N} \end{array} \right.
\ee
so that
\begin{align}\label{Stokes.div}
\int_{\mathscr{N}} (\nab{a} v^a) \, \veps & = \int_\Sigma (\tilde{n}_a v^a) \hat{\veps} + \int_{\Sigma'} (\tilde{n}_a v^a) \hat{\veps} + \nonumber\\
 & + \int_S (\tilde{n}_a v^a) \hat{\veps} + \int_H (-\ell_a v^a) \hat{\veps} \ ,
\end{align}
where each component of the boundary $\dd\mathscr{N}$ is oriented with the corresponding induced Stokes' orientation $\hat{\veps}$. It is understood that choice of the vector field $\ell^a$ comes with ambiguity, $\ell^a \to \ell'^a = \lambda \ell^a$, for some positive real function $\lambda$, leading to redefinitions $n'^a = \lambda^{-1} n^a$ and $\hat{\veps}' = \jmath^*(i_{n'} \veps)$, but the integrand above remains unchanged, as $\ell_a v^a \, \hat{\veps} = \ell'_a v^a \, \hat{\veps}'$.

\section{A sample of NLE Lagrangians} \label{app:NLEs} 

A comprehensive list of all NLE Lagrangians proposed in the literature would be enormous and not quite enlightening. Therefore, we single out just a several most significant ones.

\begin{itemize}
\item Born--Infeld \cite{Born34,BI34},
\be
\LL^\scr{(BI)} = b^2 \left( 1 - \sqrt{1 + \frac{\FF}{2b^2} - \frac{\GG^2}{16b^4}} \right) \ ,
\ee
with the real parameter $b > 0$, corresponding to the strength of the maximal field. Experimental constraints \cite{EMY17,NAM18} for the parameter $b$ give us $b \gtrsim 10^4\,(\mathrm{GeV})^2$. Born--Infeld Lagrangian is sometimes truncated, for $\FF \gg (\GG/b)^2$, to
\be
\LL^\scr{(tBI)} = b^2 \left( 1 - \sqrt{1 + \frac{\FF}{2b^2}} \right) \ .
\ee

\item Euler--Heisenberg \cite{HE36} (see also \cite{Dunne04}), in weak field expansion
\be
\LL^\scr{(EH)} = -\frac{1}{4} \, \FF + \frac{\alpha^2}{360 m_e^4} \left( 4\FF^2 + 7\GG^2 \right) + O(\alpha^3) \ ,
\ee
where $\alpha$ is the fine-structure constant and $m_e$ is the mass of the electron.

\item Ay\'on-Beato--Garc\'ia \cite{ABG98,ABG00},
\be\label{ABG}
\LL^\scr{(ABG)} = \frac{3\mu}{g^3} \left( \frac{g\sqrt{2\FF}}{2 + g\sqrt{2\FF}} \right)^{\!\frac{5}{2}} \ .
\ee
It is important to stress that coupling parameters $\mu$ and $g$ are only \emph{a posteriori} identified as mass and magnetic charge for some specific solution, such as the Bardeen black hole.

\item Power-Maxwell \cite{HM07,HM08},
\be
\LL^\scr{(pM)} = C \FF^s \ ,
\ee
with some real constants $C \ne 0$ and $s \ne 0$.

\item ModMax theory \cite{BLST20,Kosyakov20},
\be
\LL^\scr{(MM)} = \frac{1}{4} \left( -\FF\cosh\gamma + \sqrt{\FF^2 + \GG^2} \sinh\gamma \right) ,
\ee
defined with one real parameter $\gamma$, is a unique class of NLE theories which is both conformally invariant (it has vanishing energy-momentum tensor) and invariant with respect to electromagnetic duality rotations \cite{GR95}. 
\end{itemize}

\begin{acknowledgments}
We would like to thank Professor Robert M.~Wald for his invaluable remarks on paper \cite{GW01} and Professor Dmitri Sorokin for drawing our attention to the novel ModMax theory. The research was supported by Croatian Science Foundation Project No.~IP-2020-02-9614.
\end{acknowledgments}

\bibliography{nlebhth}

\end{document}